\documentclass[10pt,journal, finalsubmission, a4paper ]{IEEEtran}

\usepackage{amsmath}
\usepackage{tikz, rotating, cases, multirow,scalefnt }
\usepackage{amssymb}
\usepackage{color, array}
\usepackage{mathbbol}

\usepackage{times}
\usepackage{graphicx} % more modern
\usepackage{subfigure} 

\usepackage{algorithm}
\usepackage{algorithmic}
  
\usepackage[pdftex]{hyperref}
\def\imagetop#1{\vtop{\null\hbox{#1}}}
\usetikzlibrary{arrows}
\tikzset{>=stealth'}
\tikzstyle{edge}  =[draw,-]

\def\secref{Section~\ref}
\def\figref{Figure~\ref}

\def\eqref{Equation~\ref}

\def\d{,\ldots,}
\def\wx{\boldsymbol{w}_{\mathrm{x}}}

\def\gp{\boldsymbol{\theta}}

\DeclareMathOperator*{\argmax}{arg\,max}
\def\nu#1#2{{#1}\kern.166em\textrm{#2}}% number + unit (with small space in between)

\hypersetup{%
pdftitle={Modelling Noise-Resilient Single-Switch Scanning Systems},
pdfauthor={Your Authors},
pdfkeywords={your keywords},
bookmarksnumbered,
pdfstartview={FitH},
colorlinks,
citecolor=black,
filecolor=black,
linkcolor=black,
urlcolor=black,
breaklinks=true,
}

\begin{document}

%\title{Ticker: An Adaptive Text-Entry Method for Single-Switch, Visually Impaired Users}
\title{Modelling Noise-Resilient Single-Switch Scanning Systems}

\author{

\begin{tabular}{cccc}
Emli-Mari Nel &   Per Ola Kristensson  & David~J.~C.~MacKay \\
University of Cambridge &    University of Cambridge   &    University of Cambridge \\
en256@cam.ac.uk &  pok21@cam.ac.uk   & djcm1@cam.ac.uk \\
\end{tabular}
}

\maketitle

\begin{abstract}
Single-switch scanning systems allow non-speaking individuals with motor disabilities to communicate by triggering a single switch (e.g., raising an eye brow). 
A problem with current single-switch scanning systems is that while they result in reasonable performance in noiseless conditions, for instance via simulation or tests with able-bodied users, they fail to accurately model the noise sources that are introduced when a non-speaking individual with motor disabilities is triggering the switch in a realistic use context.
To help assist the development of more noise-resilient single-switch scanning systems we have developed a mathematical model of scanning systems which incorporates extensive noise modelling.
Our model includes an improvement to the standard scanning method, which we call fast-scan, which we show via simulation can be more suitable for certain users of scanning systems.
\end{abstract}

%  Modelling accuracy and robustness has been validated by field experts, our own software implementation, and an audio pilot study.
%  This modelling effort provides a more nuanced understanding of single-switch scanning systems' actual performance with the intended end-users and %can serve as a mathematical foundation for more sophisticated noise-tolerant systems.
  
{ \IEEEkeywords single-switch scanning systems; accessibility; augmentative and alternative communication; text entry}

\section{Introduction}

\label{sec:intro} 

 Single-switch scanning systems are a class of augmentative and alternative communication (AAC) devices.
A {\em single-switch user} is someone whose 
primary means of communication relies on toggling a switch. 
Single-switch users include non-speaking individuals with motor disabilities who are mostly confined to wheel chairs, such as users with cerebral palsy or locked-in syndrome.
 Examples of triggers for single-switch systems include blinking, raising an eye brow, or thinking of an activity such as tennis~\cite{Grafton2010, Grauman2003, Tan2010}.

\begin{figure}[!!!!!htb]
  \centering
  \begin{tabular}{|c|} \hline 
  \begin{minipage}{7.6cm} \vspace{1mm} \includegraphics[width=7.6cm]{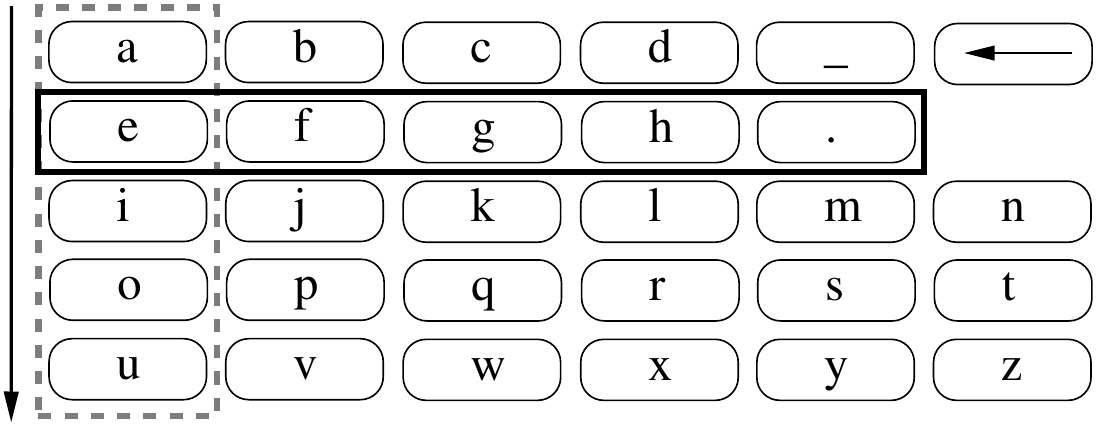}  \end{minipage}  
  \\ (a) 
  \\ \begin{minipage}{7.6cm} \includegraphics[width=6.4cm]{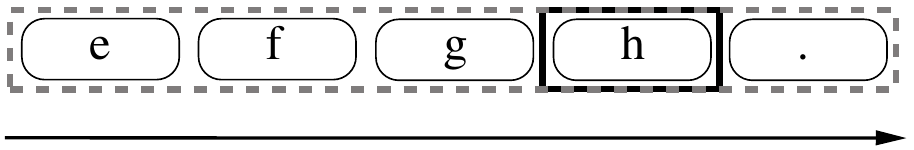}  \end{minipage} 
  \\ (b)
  \\ \hline
 \begin{tabular}{|c|c|c|} \hline 
  \begin{minipage}{2.3cm} \vspace{1mm}  \centering \includegraphics[width=2.3cm]{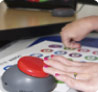}  \end{minipage} 
  &
  \begin{minipage}{2.3cm} \centering  \includegraphics[width=2.3cm]{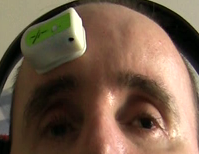}  \end{minipage} 
  &
  \begin{minipage}{2.3cm} \centering  \includegraphics[width=2.3cm]{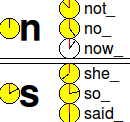}  \end{minipage}  
  \\ (c) & (d) & (e) 
  \\ \hline
    \end{tabular}

 \end{tabular}
  \caption{A typical scanning interface.    
  To select a letter, at least two clicks are necessary. In the first phase all rows are scanned in a sequence. The first click selects the desired row ``e f g h .'' (a). Thereafter the individual letter keys of the selected row are scanned in a sequence. The second click selects the desired letter ``h'' (b).
(c)-(d) Examples of switches that can be used to interface with a computer. 
  An eye-brow raise in (d) corresponds to a switch                 event, as detected
 by the shown Impulse EMG bluetooth access switch. (e) An example image of Nomon, a text entry method that can be controlled by a single switch. 
}
  \label{fig:grid2}
\end{figure}

 Scanning systems are the most prevalent text entry methods for single-switch users.
  It is an active area of research, with for instance a recent study published in~\cite{Koester2014} and new approaches being explored, such as Huffman scanning \cite{roark_huffman}. 
  \emph{Grid2} is a typical commercial scanning system widely used in practice.
 An example of how to select a letter with such a scanning system is provided in \figref{fig:grid2}.
 
The following three metrics are typically used to measure performance and they will be used throughout this article. First, text entry rate is measured in words per minute ({\em wpm}), with a word defined as five characters, including space.
Second, the number of clicks per character ({\em cpc}). Third, character error rate ({\em cer}) is defined as the minimum edit distance between between
the output word and the intentional (ground-truth) word, divided by
the number of characters in the intentional word.  
 
Although a vast number of error correction modes are possible,  the literature taking noise into account as part of the scanning-system design is rare.  
 Text entry methods are usually tested only on able-bodied users, seeking the fastest text entry rate 
 with a reasonable word or character error rate. Although it is useful to determine performance boundaries, 
it is not always clear how such empirical results would change in the presence of inevitable noise in a realistic use-context with a non-speaking individual with motor disabilities triggering a switch prone to false activations and drift.   
In noisy situations,  many impaired users are currently left with no automatic means of communication, even when having full cognitive capacity.   
 It is noted in~\cite{Koester2014} that at least \nu{6}{wpm} can be expected for a non-impaired user, whereas 
 much lower rates are  the norm for non-speaking individuals with motor disabilities (\nu{1}{wpm} or lower is common).

In this article we present the design of a plausible model of single-switch scanning systems within a probabilistic framework. Each noise source is represented by a probability distribution, which 
 can be easily estimated through a few measurements before performing the simulation. A Markov chain is used to model all 
  possible user interactions with the scanning system. A particularly useful and critical aspect of our work is the model's ability to reflect how the user would react with the system while writing a word, highlighting potential shortcomings in a way that can be easily interpreted by a non-expert. In noisy situations this is especially useful, as it can reduce the immense effort to evaluate an interface with a large representative sample of impaired users. Indeed, since the performance experienced by the end-users of single-switch scanning systems is varying due to many factors, such as the type of disability, noise characteristics of the switch, motor control ability, cognitive capacity and level of literacy, it is often impossible to evaluate such methods reliable via A/B testing in controlled experiments. We instead argue it is possible to gain design insight by using a combination of probabilistic modelling and model validation with non-speaking individuals with motor disabilities.

We conjecture that unaccounted noise sources are the main reason why text entry rates drop substantially when single-switch scanning systems are used by non-speaking individuals with motor disabilities. Rigorous noise models that provide an accurate  reflection of realistic use-contexts is the first step  to improve the robustness of the underlying design of scanning systems. By modelling the noise sources one  can potentially reduce the number of error correction substantially, which can help the users obtain text entry rates closer to the performance of able-bodied users.     

Our simulation results indicate that scanning systems are indeed very sensitive to false positives, making it difficult  to resolve by manual error correction actions.  
It is shown that the best way to deal with click-timing noise is to increase the scanning delay, which leads to reduced text-entry rates.

\section{Noise Sources}

We define the most important noise sources as follows: {\em Switch noise} comprises false acceptances and false rejections.
{\em False acceptances}, also referred to as {\em false positives}, are
spurious detections.   
 {\em False rejections}, also referred to as {\em false negatives} are switch events that are erroneously ignored.
 The  {\em click-timing noise}  causes the observed click time of a switch event to be 
earlier/later than intended. This can be due to the user's response time or due to the delay caused by
 the device that captures a switch event.  In practice the latency can easily be as much as  
 2 seconds, and sometimes much longer, depending on the complexity of the switch and the user's disability. Able-bodied adults typically exhibit about \nu{50}{ms} reaction delay when using a standard keyboard.
    
 An important practical consideration regarding the click-timing noise is a large but consistent (small variance) latency. This problem is exacerbated in an audio-based system (as opposed to a visually-based system), or when the user has a severe impairment.
If a scanning delay is increased according to the average click-time latency, it can be detrimental
for the text entry rate.

In scanning systems, sensitivity to the click-timing noise is typically reduced by increasing  the scanning delay.
False positives are typically dealt with in three ways~\cite{grid2}:
First, the duration between accepted clicks is restricted, effectively 
 modelling the recovery time of the user.  
 Thus, if, for example a whole series of clicks is received in rapid succession, 
 only the one click within the recovery-time window will be accepted. This compensates for situations  
where e.g., a faulty cable generates a series of clicks when only one click was intended. 
 Second, an undo time window 
 can be used to undo a false positive during a row scan.
 If $U$ column group scans have passed, the last row selection will be cancelled, 
and the system will resume by scanning the rows, where a {\em group scan} refers to all the scans associated with 
a specific row/column. Thirdly, a ``delete" symbol can be included in the layout to remove the last letter selection.

 \section{Background}
 \label{sec:literature}

Shannon's noisy-channel coding theorem~\cite{shannon48}
is the foundation for an information theoretical analysis of reliable communication over a noisy channel, in this case single-switch scanning systems \cite{MacKayButtons2004,
MacKay2006, Mead2007}. 
The noisy-channel coding theorem states that for any given degree of noise contamination of a communication channel, 
it is possible to communicate discrete data nearly error-free up to a computable maximum rate through 
the channel. The computable maximum rate of the channel is called the {\em channel capacity}.  
For our application, the information rate~$B$ is measured in bits per second:
\begin{equation} \label{eq:channel_capacity}
B = \frac{\mathcal{I}(\mathbf{x} ; \mathbf{y})}{T_{Y}} 
       = \frac{\mathcal{H}(\mathbf{x}) - \mathcal{H}(\mathbf{x} \mid \mathbf{y})}
         {T_{Y}}, 
\end{equation} 
where $\mathbf{x}$ is the input set (a list of words that the user intends to write), $\mathbf{y}$ is the output set (the list of words the user writes), $\mathcal{I}(\cdot)$ 
is the mutual information, 
$\mathcal{H}(\cdot)$ refers to the entropy function, e.g.,  $\mathcal{H}(\mathbf{x}) = \displaystyle \sum_{x \in \mathbf{x}} P(x) \log_2 \frac{1}{P(x)}$, and 
$T_{Y}$ is the average time it takes to produce an element in $\mathbf{y}$. $\mathcal{I}(\mathbf{x} ; \mathbf{y})$ 
measures  the degree of uncertainty in $\mathbf{x}$ after observing $\mathbf{y}$. 
If all elements in $\mathbf{x}$ can always be inferred from $\mathbf{y}$, the probability 
of error is zero and will cause the conditional entropy term 
$\mathcal{H}(\mathbf{x} \mid \mathbf{y} )$ to vanish, so that $\displaystyle B=\frac{\mathcal{H}(\mathbf{x})}
         {T_{Y}} $. 

 If $B$ is maximised with respect to $P(\mathbf{x})$, the corresponding $B$ is equal to the 
channel capacity. For example, in some problems
the channel capacity can be reached with zero probability of error by using only some of the input symbols, i.e., setting some 
values of $P(\mathbf{x})$ to zero~\cite{MacKay03}.  

 MacKay et al.~\cite{MacKayButtons2004} model a single-switch user 
as follows: A selection (letter/word/sentence) is made
 from one click to the next. The probability that the user waits for a time $D + ng$ seconds from one click to the next 
 is
\begin{equation}
p_{n} = (1-\beta)^{n} \beta, n \in \{0,1,\ldots\},
\end{equation} 
  resulting in an information rate of 
\begin{equation}\label{eq:single_switch_info}
\displaystyle B=\frac{\mathcal{H}_{2}(\beta)}{\beta D + g},
\end{equation}   
where $D$ is the reaction delay, and it is assumed that the user clicks within $g/2$ seconds around each intended click time.
The capacity is then computed by optimising $B$ with respect to $\beta$.
If the switch is unreliable (i.e., 
yielding false positives or false negatives) a fraction $f$ of the time 
it constitutes a bit error, so that the capacity will drop with a factor $1-\mathcal{H}_{2}(f)$, 
where $\mathcal{H}_{2}(f)$ is the binary entropy function (see the second part of the noisy-channel coding theorem 
described in~\cite{shannon48,MacKay03}). 
 
A low error rate is  the most important requirement for our application, even more so than the speed at which the user
can write.
A reduction in the error rate typically requires 
longer output codes (elements in $\mathbf{y}$) which increases $T_{Y}$ and reduces $B$. This doesn't have to be the 
case, shown with an example in \secref{sec:improving_grid}.

One way to increase the length of the output codes to accommodate noise (reducing the error rate) is to 
allow multiple clicks to make a selection. This effectively increases the length of the output code.
After each click, the previous click information can be used to 
update the way the input symbols are presented to the user (referred to as {\em dynamic updating}) and this can be of huge potential benefit if it can be applied. 
It is, however, crucial for a single-switch system to use as few clicks as possible for a selection. 
One can imagine how exhausting an impaired user
might find it to use more than two facial gestures to select only one letter at a time. Selecting 
parts of sentences instead of one letter at a time is therefore a highly desirable characteristic.

\eqref{eq:channel_capacity} is used as basis to further discuss existing methods. However, we follow the literature when making direct performance measurements, namely the 
text-entry rate (wpm), click rate (cpc) and character error rate (cer), as mentioned in \secref{sec:intro}.
 Many papers do not publish the last two quantities. So, following existing literature, the text entry rates of some techniques are compared, and  where possible we also mention the  click- and error rates.  

\section{Scanning Systems} 
\label{sec:scanning}

Scanning systems are widely used. Research on scanning systems is therefore still active, with a recent study published by Koester and Simpson~\cite{Koester2014}  in 2014. The latter paper focuses on case studies of impaired users, and also contains a summary of previous work in the field. Some of their findings are summarised below.  
 
 The  aforementioned study by Koester and Simpson~\cite{Koester2014}  note that a very fast user may achieve \nu{7--8}{wpm} using a single-switch scanning system, but 
 rates of \nu{1}{wpm} and lower are common. They note that at least \nu{6}{wpm} can be expected for an able-bodied user, whereas 
 much lower rates are the norm for non-speaking individuals with motor disabilities. They focussed on calculating the settings for any scanning system that 
  best suits a user's abilities. Their study included nine participants whose text entry range improved from \nu{0.3--2.9}{wpm}
  to \nu{1.1--6.5}{wpm}, with the most important determining parameters being the scanning delay, the presence of word-predictions, and the alphabet layout.
  Most users were diagnosed with cerebral palsy. It was not clear if a large variety of switches were used, as some switches may be 
  more error prone that others.
 
  If the scanning delay is too short, error corrections will have a detrimental effect on the text entry rate, 
  whereas if it is too long, unnecessary time will be wasted with each scan, also causing the text entry rate to be low.
   Koester and Simpson~\cite{Koester2014}  found that a large improvement was achieved by measuring the user's average latency and setting the scanning delay accordingly. 
   Each user was allowed to use his/her system of choice, which included a variety of commercially available systems.  Their results
  were in line with previous published results, ranging from \nu{0.5--4}{wpm} for motor-impaired users, even though they were not using their own systems. 
 
 Word predictions can increase the text entry rate of a scanning system  in some circumstances. It can also reduce the 
average number of clicks per character significantly~\cite{Koester1996, Lesher1998}.   
 Stephen Hawking's text entry 
rate doubled from \nu{1}{wpm} with word-predictions from a customised language model~\cite{hawking2014}. 

A study by Koester and Levine in 1996~\cite{Koester1996} compared the text entry rate of eight able-bodied users
to that of six users who had high-level spinal cord injuries. The able-bodied users typed using their usual method of keyboard access,
 whereas the injured users used mouth stick typing. They found that word predictions had a negative impact in this study: the use of word prediction caused the text entry rate to decrease for the spinal-cord injured users and only modestly enhanced it for the able-bodied users.
A more recent study by Koester and Simpson~\cite{Koester2014} (2014) reported an average text entry rate with/without word-predictions of \nu{2.7/1.5}{wpm}. They note, however,
that the word-prediction settings generally had to be tailored to the user's needs, which may explain the discrepancy in the literature,
where it is often reported that word-predictions have negative or no impact on the text entry rate. 

In the scanning system Grid2~\cite{grid2}, it is recommended that the word-prediction functionality is disabled  
 when using a scanning system in audio mode, or when the scanning delay tends to be long.
 From Koester and  Simpson's work~\cite{Koester2014}, we can expect an average of \nu{1.5}{wpm} (or much less)  when 
  a scanning system is used by a non-speaking individual with motor disabilities, and/or in audio mode. Also, without word predictions, two clicks are always necessary to select a character.  
 
In scanning systems, sensitivity to the click-timing noise is typically reduced by increasing  the scanning delay.  An increase to
  $k \cdot T_{\mathrm{S}}$, where  $T_{\mathrm{S}}$ is 
the scanning delay, for e.g., row $v$, will reduce the capacity by a factor $vk$ for that selection, 
where $v = \in \{1, \ldots, V\}$.

The model for scanning systems defined by Koester et al.~\cite{Simpson2011} is similar 
in spirit to ours, but doesn't allow  modelling of sequences of actions. Quoting them, they assume that:
{\em  ``the user never makes two mistakes in the same
selection attempt (i.e., the user never selects the
wrong row and then selects the wrong column; the
user never selects two incorrect rows). The model
could be expanded to accommodate this by adding in
additional probabilities for two error sequences, but it
would get very complicated." } On a theoretical level, this assumption is only valid for a very small probability of error.
We allow modelling of a large number of user actions; up to a configurable large limit. 
 
 \subsection{Dasher}
 \label{sec:dasher}
 One of the single-switch algorithms that theoretically performs the best (according to \eqref{eq:single_switch_info}) is Dasher~\cite{MacKayButtons2004,
 MacKay2006, Mead2007}. 
  Generally, Dasher depends on good vision. An untested audio version in two-button mode is available, but has  not been adapted to cope with unreliable switches~\cite{MacKayButtons2004}.  
   Dasher was originally designed as a hands-free text entry method, controlled by any continuous motion such as eye gaze, or finger movements~\cite{Ward2000}.  
  The problem is divided into two independent parts:
  \begin{enumerate}
  \item Efficient information capture by the system, i.e., optimising 
  \eqref{eq:single_switch_info}.
  \item Efficient language compression. The language model used by Dasher typically compresses English 
  to about 2 bits per character.
  \end{enumerate}
 
  Unlike most other
  techniques, a click hardly ever  maps to a single character, but   to parts of sentences.
  This allows  for fewer gestures to be used compared to existing techniques (a major advantage for physically impaired users). 
  More probable letters occupy more screen space and are therefore easier to find and quicker to select. The latter allocation space is adapted dynamically, as determined by the language model and according to previous selections made by the user. However, the layout remains fixed (letters are always listed in alphabetical order), so that the dynamic updating places no additional cognitive burden on the user.  
  
   A text entry rate of about 34 wpm is possible when using a computer mouse to control Dasher in continuous mode by an expert user.
 This result is impressive considering the typical ten-finger typing rate of 40-60 wpm~\cite{Ward2000}.  
   Extensive tests have shown that 14 wpm are possible with relatively little practise and  when using gaze as controlling device~\cite{Rough2014}.
   Depending on the individual, it has been shown that an expert user can reach up to 25 wpm  using gaze as controlling device~\cite{wardmackay2002}.  
 
   In one-button mode, an expert user has shown 
 to be able to communicate at 10 wpm using only 0.4 gestures per character~\cite{MacKay2006}. 
  Finding a way to use more screen space to display the selection options in 2D, or improving the language model, can  each lead to a possible factor 2 speedup.

 \subsection{Nomon}
 \label{sec:nomon}
  
  Nomon~\cite{nomon,nomon2} is a single-switch system that applies Shannon's noisy coding theorem in its design.  The capacity of the channel is computed in~\cite{nomon2}. Several clocks are presented on a display screen. 
There is one clock associated with each letter in the alphabet, and some additional clocks are associated with likely words at the time. 
 To make a selection, the user has to press a button when the rotating hand of the corresponding clock reaches noon. 
 Words/letters that the system thinks are probable are shown in yellow (e.g, ``not\_''), whereas improbable 
 ones are shown in white (e.g., ``now\_''), where ``\_'' is used throughout this paper to indicate a space. Several clicks may be necessary per selection, but it has been measured that, on average, two clicks are usually enough. \figref{fig:grid2}(e) depict some of the clocks.

 After each click, all possible random codes are decoded and mapped back to words. If the most probable word is above a certain threshold, it is selected. 
   Nomon  may require more than 2 cpc in the presence of  noise. Dynamic updating is done by  updating the most probable word-selection list after each letter selection. This noise-coping strategy can be compared to a scanning system where the layout changes dynamically, but without the additional cognitive load that the scanning system will impose.

 Nomon is sensitive to the latency of the user/system as all the available click-time information is needed after each click before the program can proceed.  Nomon is, however, less sensitive to the latter latency  compared to scanning systems, where the scanning delay has to be adjusted according to the latency for {\em every} scan. 
   False acceptances and rejections are not accounted for by Nomon. Like Dasher and standard scanning systems, these errors will therefore reduce the capacity by $1-\mathcal{H}_{2}(f)$. 
  
  An average of 
 1.5--2 cpc is  typically necessary to select a word, as measured empirically on able-bodied users. Using Nomon, an expert able-bodied single-switch user can write at a speed of approximately 10 wpm with 
 a low probability of error. An extensive experiment on able-bodied novice users has shown Nomon's superiority to standard hierarchical scanning systems: at the end of the experiment
 the average text entry rate for Nomon was 5.8 wpm and 4.3 wpm with the hierarchical scanning system.   

  Nomon was tested informally for our research purposes. Testing was done with a customised webcam-based switch controlled by smiling and winking. Eight able-bodied participants were asked to write
   ten phrases. 
   None of the participants were familiar with using facial gestures to communicate. After five phrases they were required to take a break (or after one hour if they wrote less than five phrases). 
   Each user trial took between 1.5 and 2 hours. 
   The users were allowed to increase the speed at any time, but the speed was not increased up to breaking point. 
   The participants were requested to write as accurately as possible (at a comfortable speed), instead of writing as fast as possible.  
   The main goal was to test our customised 
   gesture switch (which has a low error rate), to see if it could be used to communicate. A secondary goals was to
   determine performance estimates if Nomon is used by controlling it with a gesture switch instead of a joystick button. 
  
   All participants  were able to write comfortably at speeds between 1.5 wpm and 2 wpm with very low probability of error, averaging approximately 2 cpc. 
   Note that 
  the scanning delay in Nomon had to be slowed down with as much as 800 ms in some cases (some users can easily take up to one second to make a smiling gesture). 
   When the user can click extremely precisely (if the click-timing distribution is close to the 
  delta function), as little as one click per selection may be necessary. Although this is an informal experiment, it provides a strong indication that 
  Nomon is sensitive to even a response latency such as \nu{800}{ms} as expected (mentioned above).
 
 % POK: I commented out the paragraph below; it does not appear to be that relevant to the manuscript's objectives
%  Theoretically, Dasher is superior to Nomon as a single-switch text entry method---frequently occurring sentences can be selected much quicker with less clicks. Some contributing factors are: more efficient language model integration in Dasher, 
 % and the information loss due to the  periodicity of the clocks in Nomon~\cite{nomon2}. Consider, however,  
%  making graphical user interface (GUI) selections such as switching on the television, calling for help, or opening a file. Typically, these selections do not  rely on a language model (as each selection is equally likely). Nomon is more suitable than Dasher for such selections, as the corresponding clocks can be easily overlayed with the Desktop, making them quicker and easier to find than in Dasher. 

\section{A Model of Noise-Resilient Scanning}
 
To do a performance analysis, expectations regarding the text entry rate ($\# scans$), the number of clicks ($\# clicks$), and the number of character errors ($\#errors$) are computed, where $\# scans$ is multiplied by an appropriate scan delay where necessary to covert it to $wpm$.   
As a performance metric, we compare the first and second-order statistics of the latter entities.  
  Probability mass functions of the quantities in question are derived numerically, so that expectations can be computed.

In all simulations,  a phrase is processed one word at a time. Each word can be correct, containing spurious characters, or a time out error can occur if the user takes excessively long. Expectations are computed by averaging results over all processed words.

 We make use of the recommended
settings in a popular and representative hierarchical scanning system~Grid2\cite{grid2}. The settings are tailor-made for
  impaired users (who have latencies of greater than \nu{1}{second}). The simulation for this user group (who probably needs it most) is an accurate reflection of the real world. 
 We therefore do not include all error correction modes, e.g., we do not allow for reverse scanning and word predictions. More latent states can be added to the Markov chain (which is used to model user actions) to include more modes. 
 
Spurious row scans are assumed to be corrected through an undo time window, whereas spurious column scans are assumed to be corrected 
with a delete symbol in the layout. We do not model the effect of limiting the duration between accepted clicks, 
  although this can also be easily simulated, if necessary. The user is assumed to immediately follow the action to correct any errors, which can result in a correction or a new sequence of 
errors to be corrected.  
 
To model all possible outcomes when interacting with a scanning system is not immediately obvious, as there is potentially an infinite number of possibilities. We therefore define certain system-failure conditions to decrease the number of possibilities. System failure represents a user who gives up on the system as it becomes unusable.

   For validation purposes we have tested the free trial-version Grid2 software. We have encountered some problems, specifically related to using it in audio mode.
 Firstly we have found it difficult to select the first cell of a group scan at a high speed as it is difficult to anticipate when to click.
  In such cases, it is recommended to increase the scanning delay of only the first scan.   However, in practice, it then becomes more difficult to 
  time the other scans. We also found the software  difficult to use  if the scanning delay is shorter than \nu{1}{second} in audio mode (probably due to some software implementation issues).
  
 For the reasons above, we have implemented our own scanning system to facilitate usage in audio-mode. The code can be downloaded, and may also be useful for the reader to better envision the usage resembled by the simulations. It was also used to validate our simulation results by conducting a pilot study. 
 
 In our software, all sound files were overlayed with a soft ``tick'' sound, marking the beginning of each cell. An additional ``tick'' sound has also been included before the first letter of each group scan.
 We have  used our own sound files which are faster recordings of the alphabet than the default sound files provided by the Grid2 software. With these small modifications, one can reduce the scanning delay to a mere \nu{300}{ms}, which reflects a fast scanning delay when used in visual mode (see e.g.,~\cite{nomon}). The ``tick'' sound provides a rhythm when used in audio mode, that can help the user tremendously to anticipate  when to click, especially if the user has memorised the layout. Although an extra ``tick'' cell is included per  row/column group, the overall speed is increased. Our software also counts the number of scans used to write a word, which is useful for performance comparisons.

\subsection{Markov Chain}
 
Let $V$ be the maximum number of scans that constitutes a group scan, i.e., $V=\max(I,J)$, where $I$ is the visible number 
of rows in the layout, and $J$ is the visible number of columns. The sound file of the first visible cell is seen as a continuation of the ``tick'' sound that precedes it in audio mode. 
The scanning delay of each cell is $T_{\mathrm{S}}$ (measured in seconds), except for the first cell which has a scanning delay of $2T_{\mathrm{S}}$.

 The user's intentions are modelled with the random variable $v \in  \{\mathbf{v}, \emptyset\}$, where $\mathbf{v}= \{1, \ldots, V\}$: If the user intends to make a selection, $v \in  \mathbf{v}$. Otherwise, if the 
 user waits for the undo time window to pass $v=\emptyset$. 
 
 Initially it is a assumed that the user's click-timing is well represented by a Gaussian distribution. In later sections, however, we show that 
the derivations to follow apply equally will with any other distribution that is continuous in time. 
 
If the user's intention is to select cell~$v$, the click-timing~$t$ is modelled by  $t \sim \mathcal{N}(\mu_v,
  \sigma)$, where $\mu_v =  \mu_v^{\mathrm{B}} +0.5\cdot T_{\mathrm{S}}$, 
  where
   $\mu_{v}^{\mathrm{B}} \in [T_{\mathrm{S}}, V \cdot T_{\mathrm{S}}]$ is the starting time of cell~$v$. 
   The beginning of a group scan (after the ``tick" sound) is therefore $\mu_{1}^{\mathrm{B}} = T_{\mathrm{S}}$.

The Gaussian mean is therefore aligned with the centre of each cell's scanning-window time if $\Delta=0$.  
 The latency $\Delta$ and the standard deviation  $\sigma$  of the Gaussian distribution are assumed to be the same for all cells.   Note that, although the Gaussian-noise assumption simplifies the problem somewhat, $\sigma$   provides much more information regarding the 
click-timing noise than only the average response times which was used in~\cite{Koester2014} to adapt the user settings,
and can be measured as easily.

 An important consideration is how to model false positives.
 Consider the practical example, where an Impulse EMG switch generates false positives when the user's body temperature rises. 
 If the scanning system is used for 5 minutes, one would expect more false positives than when using it for 30 seconds.   
 A noise model that can take the false positive {\em rate} into account is the homogeneous Poisson process~\cite{kingman-poisson-processes}.
  %A Poisson distribution arises when the number of trials in the Binomial distribution goes to infinity and the probability of the binary event under question is very small. 
 Following, the number of events (spurious clicks) $N$ in a finite time interval $T$, 
 always has a Poisson distribution, i.e., $P( N \mid T, \lambda) =  \frac{(\lambda T)^{N} e^{-\lambda T} }{N!}$, 
 where $\lambda$ is the average number of false positives per unit time. Hence, the longer one waits, the 
 smaller the probability that $N=0$, which is an accurate and important reflection of the reality for our application. 
 
% If all $N$ click times in a set $\{t_{1}, \ldots, t_{N}\}$, are generated by a Poisson process,  
% $P( \mathbf{t} \mid N, \lambda, T) = \frac{1}{Z_{\mathrm{N}}} \left(\frac{1}{T}\right)^{N}$,  where $t_{1} < t_{2} < \ldots < t_{N}$ and $Z_{\mathrm%{N}} = \int_{0}^{T} \int_{t_{1}}^{T} \ldots \int_{t_{N}}^{T} \frac{1}{T^{N}} dt_{N}, \ldots, dt_{2} dt_{1} = \frac{1}{N!}$.
% It follows that: 
%\begin{equation} \label{eq:poisson}
% P(\mathbf{t}, N \mid \lambda, T )
% =  P(\mathbf{t} \mid N, T ) P(N \mid \lambda, T) 
% = \lambda^{N} e^{-\lambda T}.      
%\end{equation}    

False negatives are represented by the Bernoulli distribution: The probability that a switch-event is erroneously ignored is $f$. The switch noise parameters, $\lambda$ and $f$, can be measured in practice, and are in many cases, part of the switch specifications.
All noise sources are assumed to operate independently from each other.

It follows that the probability to receive no clicks ($c=0$) while aiming for $v$ and scanning cell~$v'\in  \mathbf{v}$ is:
 \begin{align} \label{eq:p0}
 p_{0}(v', v) &= P(c=0 \mid v',v) \\ 
              &= e^{-\lambda T_\mathrm{S}}  \left[ 1  - (1-f)q(v', v) \right]^{\mathbb{1}[v \neq \emptyset]},
 \end{align}   
 where
 \begin{equation}\label{eq:overlap}
q(v', v) =  \int_{\mu_{v'}^{\mathrm{B}}}^{\mu_{v'}^{\mathrm{B}} + T_\mathrm{S}}  \mathcal{N}(t \mid \mu_{v}, \sigma) dt,
\end{equation} 
 $\mathbb{1}[\cdot]$ is the indicator function.
 
It follows that the probability to receive  a click
in the same scenario is: 
 \begin{equation} \label{eq:p1}
 p_{1}(v', v) =  P(c = 1 \mid v',v) = 1 -  p_{0}(v', v). 
 \end{equation}

The distributions for false- positives and negatives are generic and will probably be suitable for most switches (switch error rates are typically defined in their user manuals). However, the click-timing distribution is expected to vary between users.
For example,  the Gaussian can be replaced with a mixture model, where the second component represents an unusual lag in response (e.g., to represent a state of fatigue). 
Other than statistical independence, the simulation program is invariant to the choice of noise distributions. 
Replacing the Gaussian will require only recalculating \eqref{eq:p0} and \eqref{eq:overlap}. 
With the Gaussian example, it is easy to see that the $ T_{\mathrm{S}}$ 
should probably be increased linearly with $\Delta$ and $\sigma$, otherwise the user can make  unintentional selections with a high probability 
that must be corrected afterwards

A Markov chain is constructed for each intentional word $\mathbf{w}_{\mathrm{x}}$ to represent the user actions defined in the manual (for the target group), up to the point where it becomes infeasible to use the scanning system.  
A word selection is assumed after spacebar (``\_'') or fullstop (``.'') is selected.
 The dynamic process is terminated if a word $\mathbf{w}_{\mathrm{y}}$ is selected or if the system fails.   Three terminating states are defined as such. 
 If the system fails, the terminating {\em failure state} will be reached, as the user is unable to complete a word.  The {\em correct state} is reached if $\mathbf{w}_{\mathrm{y}} = \mathbf{w}_{\mathrm{x}}$, whereas the {\em error state} is reached if $\mathbf{w}_{\mathrm{y}} \neq \mathbf{w}_{\mathrm{x}}$. The simulation is done over a finite duration
$T=\kappa\times M\times I \times J \times T_\mathrm{S}$ seconds, where $M=|\mathbf{w}_{\mathrm{x}}|$. If the user takes longer than $T$ seconds to write a word, system  failure is assumed (as a user might want to give up trying to write the word). This type of failure is called a {\em time-out} failure.

After an erroneous row selection, the user is assumed to wait for $U$ group scans to undo it. The undo time window is represented by the random variable $u$. 
During a row scan $u = \emptyset$. Otherwise, $u \in \mathbf{u}$, where $ \mathbf{u} \in \{0 \d U\}$, representing the number of column groups scans that have passed.
After an erroneous letter selection, the user is assumed to immediately proceed with an attempt to delete it. The number of spurious letter selections is  represented by $e\in \mathbf{e}$, where  $\mathbf{e}=\{0 \d E\}$. If $e=E$ the system will fail.
 
 The simulation parameters are summarised by 
\begin{equation}\label{eq:theta}
\gp = \{ 
\Delta, \sigma, f, \lambda, T_\mathrm{S}, U, E, \kappa \},
\end{equation} 
where
\begin{itemize}
\item $\Delta, \sigma$: Gaussian click-timing parameters.
\item $f$: False negative probability. 
\item $\lambda$: False positive rate (per second). 
\item $T_\mathrm{S}$:  Scanning delay (seconds).
\item $U$: Number of column scans required to trigger an undo.
\item $E$: Number of spurious letter selections before system failure. 
\item $\kappa$: $T \propto \kappa\times M$ specifies the maximum time that can be spent on a word before  system failure is assumed. 
\end{itemize}
  
\begin{figure}[!!!!!htb]
\centering
\begin{tabular}{|c|}  \hline
\begin{minipage}{2cm} \vspace{1mm} \includegraphics[width=2cm]{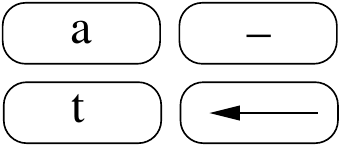} \end{minipage}
\\ \hline
\end{tabular} 
\caption{(a) A simple Grid2 $2 \times 2$ layout, where the symbol
 ``$\leftarrow$" will delete the previous output symbol. } 
\label{fig:grid_simple}
\end{figure}

\begin{figure*}[!!!!!htb]
{
\setlength{\extrarowheight}{-1mm}
\input{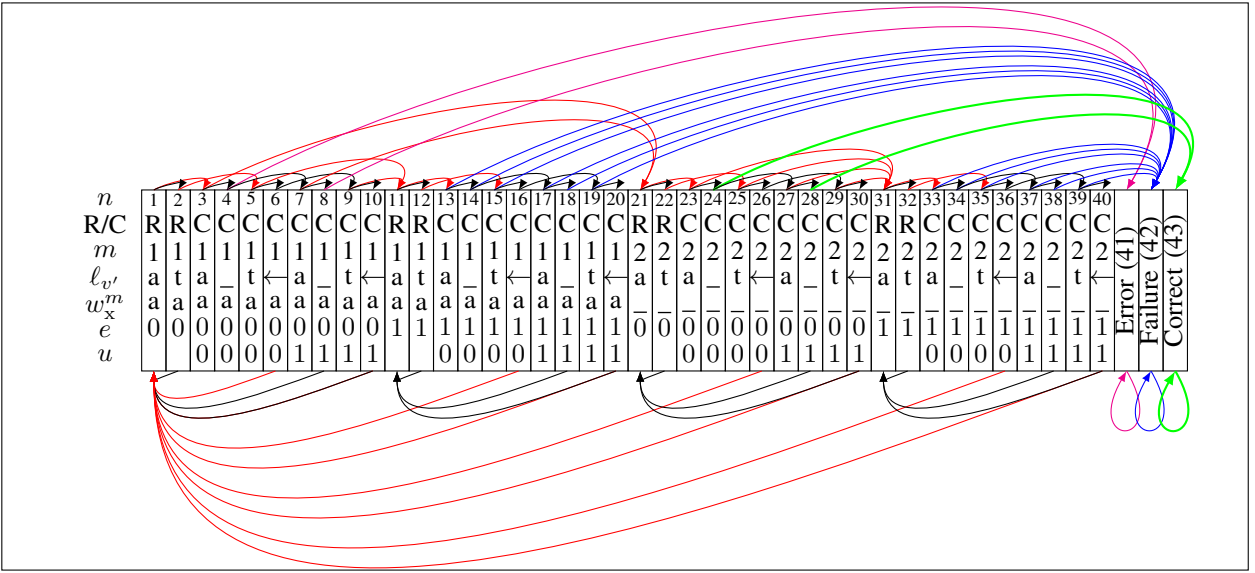}
}
\caption{A state diagram to visualise the latent states of Figure~\ref{fig:grid_simple} for $\wx=$``a\_''.  
Each state (rectangle) is associated with an $id$ (\eqref{eq:state_id}), 
where for readability, R/C corresponds to $r=1/r=0$, indicating a row/column scan.
  The transitions to the terminating correct, error, and failure states are rendered with green, pink and blue arrows, whereas all other transitions are rendered as red ({\em click}) and black ({\em miss}) arrows, with corresponding transition probabilities $a_{n'n}^{\mathrm{click}}$ and  $a_{n'n}^{\mathrm{miss}}$ (see \eqref{eq:transition_probs}). 
 }
\label{fig:grid2_wait}
\end{figure*}

A simple layout is shown in \figref{fig:grid_simple}. 
The corresponding Markov Chain for $\mathbf{w}_{\mathrm{x}}=\mathrm{``a\_"}$ is shown in \figref{fig:grid2_wait}.
Each state~$n \in  \mathbf{n}$,  where $\mathbf{n}=\{1,\ldots, N\}$, is associated with a unique $id$ ensemble:
\begin{equation}\label{eq:state_id}
 id(n) = \{r, v', m, e, u\},
\end{equation}
 where $m \in \mathbf{m}, \mathbf{m}=\{1 \d M\}$, $r \in \{0,1\}$ and $v' \in \mathbf{v}$.
The letter associated with cell~$v'$  in the layout is represented by $\ell_{v'} \in \boldsymbol{\ell}$.
 A specific intentional letter is written as  $w_{\mathrm{x}}^{m}$  (the $m$'th letter in $\mathbf{w}_{\mathrm{x}}$) and 
 $\mathbf{w}_{\mathrm{x}}^{m:m'}$   denotes the set of letters $m$ to $m'$ of the input word. For a row scan, $r=1$ and, by definition, $u = \emptyset$. For a column scan $r=0$ and $u \in \mathbf{u}$.

%\begin{landscape}

For the example, in \figref{fig:grid_simple} and \figref{fig:grid2_wait}, $M=2$, $V=2$, $U=2$, and $E=2$. 
The intentional $v$ can be deduced from $r,m,e$ and $u$. For example, state~17 represents a column scan ($r=0$) while
the intentional letter is  ``a'', and the system is busy scanning cell ``a''. One erroneous letter has been written ($e=1$), 
and one column group scan have passed ($u=1$). Since $e=1$, the user's intention is to select the delete symbol, but the wrong 
column group is currently being scanned. The user must therefore first wait for $u=2$, i.e., proceeding to state~18 and then state~11 to  first undo the erroneous row selection. If the user accidentally selects ``a'' when at state~17, two erroneous letters will be written and a transition to the system failure state~42 will be made. 
 
The last three states in the chain represent the terminating states (which have only self-loops with transition probabilities of 1.0). All other (non-terminating) states have only two possible transitions, namely one associated with a {\em click} (\eqref{eq:p1}), and the other with a {\em miss} (\eqref{eq:p0}). 
After determining the Markov chain topology, all non-zero transition probabilities can be computed from:
\begin{align} \label{eq:transition_probs}
 a_{n'n} &= P(s_{t}=n \mid s_{t-1} = n', \gp)  \\
         &= \sum_{c_{t} \in \{0,1,\emptyset\}} P(s_{t}=n, c_{t} \mid s_{t-1} = n', \gp) \nonumber  \\
         &= \begin{cases}  
       a_{n'n}^{\mathrm{click}} \, \mathrm{or}  \, a_{n'n}^{\mathrm{miss}} &  \text{if } n' < N-2, 1 \leq t < T,   \nonumber  \\
       1.0 &  \text{if } t \geq 1, n'=n, n \geq N-2,  \nonumber \\
           &  \text{or } t=T, n = N-1, n' < N-2 
 \end{cases} 
\end{align}
where $(n,n')\in \mathbf{n}$, $t \in \mathbf{t}, \mathbf{t}=\{1 \d T\}$, $s_{t}$ denotes the hidden state at time step~$t$, $c_{t}=1/0$  indicates a click/miss
that can result in a state transition (applicable to transitions from non-terminating states), whereas $c_{t}=\emptyset$ indicates that 
clicking is not possible (applicable to transitions from terminating states). Initially, $P(s_{0} = 1) = 1.0$. 

The 
click and miss probabilities, $a_{n'n}^{\mathrm{click}}$ or $a_{n'n}^{\mathrm{miss}}$, associated with non-terminating states ($n' < N-2$) can be computed from Equations~\ref{eq:p0},
 \ref{eq:p1}  and \ref{eq:state_id} (if there exists a link between states~$n'$ and $n$). More specifically, $id(n)$ is firstly computed from \eqref{eq:state_id} to determine $v$, so that $a_{n'n}^{\mathrm{miss}}=p_{0}(v', v)$  or  $a_{n'n}^{\mathrm{click}}=p_{1}(v', v)$.  Transition probabilities from terminating states are 1.0. 
 Transition probabilities from all non-terminating states at the last time step $t=T$ are given by $a_{n', N-1}=1.0$.

The probability 
of any state sequence can be computed from:
\begin{equation}\label{eq:state_seq_prob}
P(s_{0}\d s_{T} \mid \gp) = \delta(1-s_{0}) \prod_{t=1}^T a_{s_{t-1}, s_{t}}.
\end{equation}

A possible sequence for \figref{fig:grid2_wait} is: The user selects row~``a" too late, thereby accidentally selecting row~``t",
waits for the undo time window to pass, but selects column~``t" unintentionally, resulting in a spurious output symbol ``t".  
When trying to undo/delete the spurious~``t", the user then accidentally clicks on row~``a"),
 and then waits for two column scans to undo the erroneous row selection. 
 The user then manages to delete the spurious ``t" and writes ``a\_" faultlessly. 
 The corresponding state sequence is shown in the second column of Table~\ref{table:grid2_seq_example}.

\begin{table}[!!!!!htb]
\centering
{\footnotesize
\begin{tabular}{|c|c|c|c|c|c|c|c|c|c|l|}  \hline
$t$ & $n$ & $r$ &  m & $w_\mathrm{x}^{m}$  & v' & $\ell_{v'}$  & $e$        & $u$    & v 	    &   $a_{n'n}$        \\ \hline
 0  & 1   &  -  &  - &        -            &  -  & -           &  -         & -      & -  	    & 1.0           \\ \hline
 1  & 1   &  1  &  1 &        a            & 1	& a    	       &  0         & -      & 1  	    & $p_{0}(1,1)$   \\ \hline
 2  & 2   &  1  &  1 &        a            & 2 	& t    	       &  0         & -      & 1 	    & $p_{1}(2,1)$   \\ \hline
 3  & 5   &  0  &  1 &        a            & 1	& t    	       &  0         & 0      & 2  	    & $p_{1}(1,2)$   \\ \hline
 4  & 11  &  1  &  1 &        a            & 1	& t    	       &  1         & -      & 2  	    & $p_{1}(1,2)$   \\ \hline
 5  & 13  &  0  &  1 &        a            & 1	& a    	       &  1         & 0      & $\emptyset$  & $p_{0}(1,\emptyset)$   \\ \hline
 6  & 14  &  0  &  1 &        a            & 2	& \_           &  1         & 0      & $\emptyset$  & $p_{0}(2,\emptyset)$   \\ \hline
 7  & 17  &  0  &  1 &        a            & 1	& a    	       &  1         & 1      & $\emptyset$  & $p_{0}(1,\emptyset)$   \\ \hline
 8  & 18  &  0  &  1 &        a            & 2	& \_           &  1         & 1      & $\emptyset$  & $p_{0}(2,\emptyset)$   \\ \hline
 9  & 11  &  1  &  1 &        a            & 1	& a    	       &  1         & -      & 2  	    & $p_{0}(1,2)$   \\ \hline
10  & 12  &  1  &  1 &        a            & 2	& \_           &  1         & -      & 2  	    & $p_{1}(2,2)$   \\ \hline
11  & 15  &  0  &  1 &        a            & 1	& \_   	       &  1         & 0      & 2  	    & $p_{0}(1,2)$   \\ \hline
12  & 16  &  0  &  1 &        a            & 2	& $\leftarrow$ &  1         & 0      & 2  	    & $p_{1}(2,2)$   \\ \hline
13  & 1   &  1  &  1 &        a            & 1	& a    	       &  0         & -      & 1  	    & $p_{1}(1,1)$   \\ \hline
14  & 3   &  0  &  1 &        a            & 1 	& a    	       &  0         & 0      & 1  	    & $p_{1}(1,1)$   \\ \hline
15  & 21  &  1  &  2 &        \_           & 1	& a    	       &  0         & -      & 1 	    & $p_{1}(1,1)$   \\ \hline
16  & 23  &  0  &  2 &        \_           & 1 	& a    	       &  0         & 0      & 2 	    & $p_{0}(1,2)$   \\ \hline
17  & 24  &  0  &  2 &        \_           & 2 	& \_           &  0         & 0      & 2 	    & $p_{1}(2,2)$   \\ \hline
18  & 42  & -  &  - &         -            &         -        & -          &  -     &  - & - &   1.0          \\ \hline

\end{tabular}
} 
\begin{tabular}{l@{ }c@{ }ll@{ }c@{ }l}
\\
\end{tabular}
\caption{An example state sequence from the Markov chain in \figref{fig:grid2_wait}.
The state-sequence probability is the product of all the probabilities shown in the last 
column (\eqref{eq:state_seq_prob}).} 
\label{table:grid2_seq_example}
\end{table}
 
Computing $P(s_{t} = n \mid \gp)$ requires the summation over all possible state sequences:
\begin{equation}\label{eq:alpha}
\alpha_{t}(n) = P(s_{t} = n \mid \gp) = \sum_{n'=1}^{N} a_{n'n} \alpha_{t-1}(n'), 
\end{equation}
where $t \in \mathbf{t}$, $n \in \mathbf{n}$ and $\alpha_{0}(n)= \delta(n-1)$.

There is not a direct mapping between~$t$ and $\#clicks$, as the user can sometimes miss a click. 
Thus, all possible $\#clicks \in \{0 \d T-1\}$ have to be considered at each time step. Note
that $\#clicks$ can not change once a terminating state is reached.
As a first step towards computing a probability mass function for $\#clicks$, let
 \begin{align} \label{eq:click_prob_state_probs}
 \alpha^{\mathrm{clicks}}_t(r, n)
&= P(\#clicks=r, s_{t}=n \mid  \gp) \nonumber   \\     
&=  \sum_{n'=1}^{N-3} a_{n'n}^{\mathrm{clicks}} \alpha^{\mathrm{clicks}}_{t-1}(r-1,n') \mathbb{1}[t < T]   \nonumber  \\
&+  \sum_{n'=1}^{N-3} a_{n'n}^{\mathrm{miss}} \alpha^{\mathrm{clicks}}_{t-1}(r,n')\mathbb{1}[t < T]\nonumber  \\
&+  \alpha^{\mathrm{clicks}}_{t-1}(r,n) \mathbb{1}[n \geq N-2]  \nonumber \\
&+  \sum_{n'=1}^{N-3} \alpha^{\mathrm{clicks}}_{t-1}(r,n') \mathbb{1}[n=N-2, t=T],
\end{align}
where $\alpha^{\mathrm{clicks}}_0(r, n)= \mathbb{1}[r=0, n=1]$ and $r \in \{0 \d T-1\}$. Finally,
\begin{equation} \label{eq:click_probs}
P(\#clicks=r  \mid  \gp) = \sum_{n=N-2}^{N} \alpha^{\mathrm{clicks}}_T(r, n). 
\end{equation}

The probability mass function for $\#scans$ is computed in a similar way:
 \begin{align} \label{eq:scan_prob_state_probs}
 \alpha^{\mathrm{scans}}_t(r, n)
&= P(\#scans=r, s_{t}=n \mid  \gp) \nonumber   \\     
&=  \sum_{n'=1}^{N-3} a_{n'n} \alpha^{\mathrm{scans}}_{t-1}(r-r_{n'},n') \mathbb{1}[t < T]   \nonumber  \\
&+  \alpha^{\mathrm{scans}}_{t-1}(r,n) \mathbb{1}[n \geq N-2]  \nonumber \\
&+  \sum_{n'=1}^{N-3} \alpha^{\mathrm{scans}}_{t-1}(r,n') \mathbb{1}[n=N-2, t=T],
\end{align}
where $r_{n'} \in \{1,2\}$ represents the number of scans associated with state~$n$. That is, $r_{n'}  = 2$ if $n$ is associated with the first cell, otherwise $r_{n'}=1$. As an initial condition, 
$\alpha^{\mathrm{scans}}_0(r, n)= \mathbb{1}[r=2, n=1]$.
Finally,
\begin{equation} \label{eq:scan_probs}
P(\#scans=r  \mid  \gp) = \sum_{n=N-2}^{N} \alpha^{\mathrm{scans}}_T(r, n). 
\end{equation}
If $r_{n'}=1, \forall{n'}$ (i.e., ignoring the extra ``tick'' sound), then \eqref{eq:scan_probs} simplifies to:
\begin{align}\label{eq:scans}
P(\#scans = t \mid \gp)
&= P(s_{t} \geq N-2, s_{t-1} < N-2 \mid \gp) \nonumber \\  
&= \sum_{n=N-2}^{N}\sum_{n'=1}^{N-3} a_{n'n} \alpha_{t-1}(n'). 
\end{align}

To compute the number of erroneous characters, one can firstly consider the case where $\mathbf{w}_{\mathrm{x}}=\mathbf{w}_{\mathrm{y}}$: 
\begin{equation} \label{eq:no_errors}  
P(\#errors = 0 \mid \gp) =  P(s_{T} = N \mid \gp) = \alpha_{T}(N).
\end{equation}
In all other cases ($\#errors > 0$) the hidden number of correct letter selections $m_{t}$ and the hidden number of spurious clicks 
$e_{t}$ at each time step~$t$  are considered in conjunction with $c_{t} \in \{0,1,\emptyset\}$. Let
\begin{equation}
z_{t} = (m_{t}, e_{t}, c_{t}),
\end{equation} 
 where $m_{t} \in \mathbf{m}^{*}$,  $\mathbf{m}^{*}=\{0,\d M-1\}$
and  $e_{t}\in \mathbf{e}$.   
When $t < T$, a terminating state ($s_{t} \geq N-2$) can only be reached from a non-terminating state ($s_{t-1} < N-2$) if the user clicked at the previous time step ($c_{t-1}=1$). To represent termination with a spurious click, let:
\begin{align}
 & \alpha_{t}^{*}(m^{*}, e, 1)  = P(m_{t}=m^{*}, e_{t}=e, c_{t-1}=1 \mid \gp ) \nonumber \\
 &= P(m_{t-1}=m^{*}, e_{t-1}=e-1, c_{t-1}=1 \mid \gp ) \nonumber \\ 
 &=  P(z_{t-1} = (m^{*}, e-1, 1),  s_{t} \in \mathbf{n''}, s_{t-1} < N-2  \mid \gp) \nonumber \\
 &= \sum_{n\in \mathbf{n''}} \sum_{n'=1}^{N-3}   a_{n'n}^{\mathrm{click}} \alpha_{t-1}(n') \mathbb{1}[\mathbf{id}_{n'}^{3:4}=(m^{*}, e-1)],
\end{align}
where $\mathbf{n''}=\{N-2,N-1\}$ and  $\mathbf{id}_{n}^{3:4}$ represents the third and fourth elements of the $id(n)$ given by \eqref{eq:state_id}.

Once any of the terminating states have been reached, the user can not click anymore, and the number
of correct and erroneous letter selections remains fixed if the simulation is continued. To model the latter scenario, let
\begin{align}
  \alpha_{t}^{*}(m^{*}, e, \emptyset)   
&= P(m_{t}=m^{*}, e_{t}=e, c_{t-1}=\emptyset \mid \gp ) \nonumber \\ 
&= P(m_{t-1}=m^{*}, e_{t-1}=e \mid \gp ) \nonumber \\ 
&=   \alpha_{t-1}^{*}(m^{*},e)  \label{eq:chr_err_terminating_prob}, 
\end{align} 
where
\begin{align} \label{eq:chr_err_prob_t}
  \alpha_{t}^{*}(m^{*},e) 
&=   P(m_{t}=m^{*}, e_{t}=e \mid \gp) \nonumber \\
                          &= \alpha_{t}^{*}(m^{*}, e, 1)   +   \alpha_{t}^{*}(m^{*}, e, \emptyset). 
\end{align}

If the system has not terminated at the last time step~$T$, a transition to the failure state is enforced, so that:
\begin{align} 
&  \alpha_{T}^{*}(m^{*}, e)  =  P(m_{T}=m^{*}, e_{T}=e \mid \gp) \nonumber \\
&=  \alpha_{T-1}^{*}(m^{*}, e) + \sum_{n=1}^{N-3} \alpha_{T-1}(n) \mathbb{1}[\mathbf{id}_{n}^{3:4}=(m^{*}, e)]. \label{eq:last_time_step}  
\end{align}
Finally, 
\begin{equation} \label{eq:chr_err_prob}
P(\#error=k\mid \gp) = \sum_{m^{*},e} \alpha_{T}^{*}(m^{*}, e)  \mathbb{1}[k = M-m^{*}+ e], 
\end{equation}
where $k \in \{1 \d M + E\}$.

\begin{figure}[!!!!!htb]
\centering
\begin{tabular}{c}  
\begin{minipage}{5.5cm}\includegraphics[width=5.5cm]{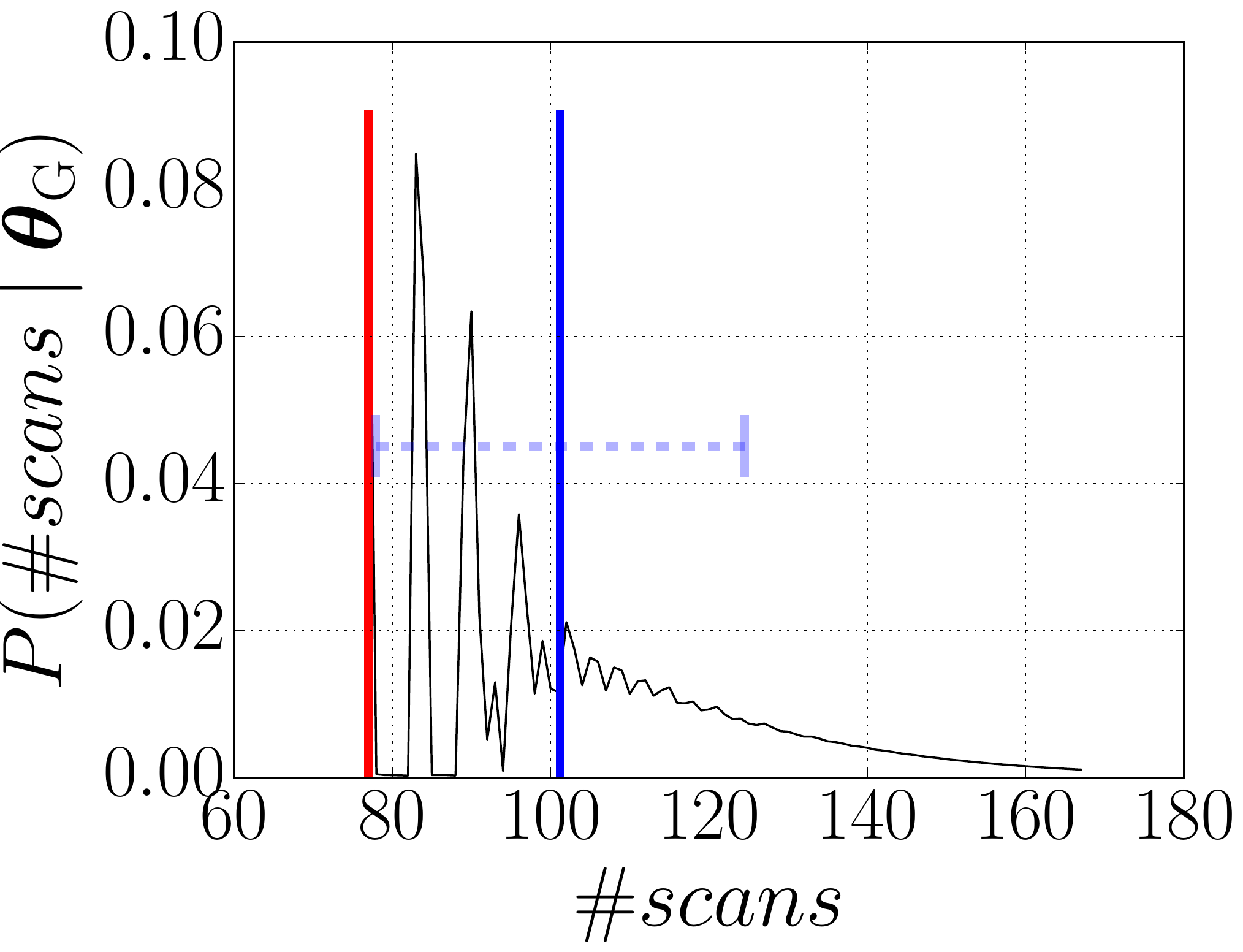} \end{minipage}
\\  
\end{tabular}
\caption{Example result  for $\wx=$``standing\_" using the configuration in  \figref{fig:grid2}. 
The distribution defined by Equations~\ref{eq:scan_probs} 
is shown (black).  The parameters (see \eqref{eq:theta}) are \nu{$\Delta=0.1$}{s}, \nu{$\sigma=0.1$}{s}, 
$f=0.1$, \nu{$\lambda=0.01$}{/s} (one false positive every two minutes on average), $E=2, U=2, \kappa=10$, \nu{$T_{\mathrm{S}}=1$}s.  
The red line (at 77 scans) indicates the best possible outcome. This can be verified by manually counting the number of scans to write the word, including  the extra ``tick" sound. The blue lines indicate  the  mean-
and standard deviation of the number of scans. }
\label{fig:grid_sentence example}
\end{figure}

Example expectations for the number of scans are given in \figref{fig:grid_sentence example}.

\subsection{Model Parameters}
\label{sec:simulation_parameters}

This section provides a motivation for the parameters that were tested.  
According to an expert field analyst~\cite{specialeffect}, it is rare that a non-speaking individual with motor disabilities can use a hierarchical scanning system with \nu{$T_{\mathrm{S}} < 1$}{s}, mainly due to the latency associated with detecting a switch event such as an eye-brow raise. It is  assumed that a novice user who can use a facial-gesture type of switch well, would start 
an average of about \nu{$T_{\mathrm{S}} = 3$}{s} (\nu{0.57}{wpm})~\cite{grid2} and progress to an average of about \nu{$T_{\mathrm{S}} = 1$}{s} (\nu{1.7}{wpm}) in noiseless conditions.

We carried out a pilot study with a single participant to determine approximate values for $T_{\mathrm{S}}$ in different conditions, and to validate our simulation model. 
The participant made use of our custom-made scanning system,  controlled by ``space bar'' on a standard keyboard. 
The participant was gradually trained to use the system blind-folded, after several hours of practise.  
Synthetic noise was added to 
the software at the end of the pilot study. 
Since the participant reached expert-level performance before adding the noise, the effect of the synthetic noise could be isolated. 
This allowed us to directly compare against our simulation results for validation purposes.  

Synthetic noise was added by sampling from the click-time- and switch-noise distributions. For example, after the user clicked, a  time delay was sampled from $\mathcal{N}(\Delta, \sigma)$, and added to the user's actual click time. Likewise, false positives were generated from a Poisson process, and each true click was accepted with probability $1-f$ or rejected with probability $f$.

\begin{figure}[!!!!htb]
\begin{tabular}{c@{}c@{}}

\begin{tabular}{c@{}c@{}c@{}}
\hspace{-0.5cm}
\begin{minipage}{1.35cm} \centering \includegraphics[height=5.6cm]{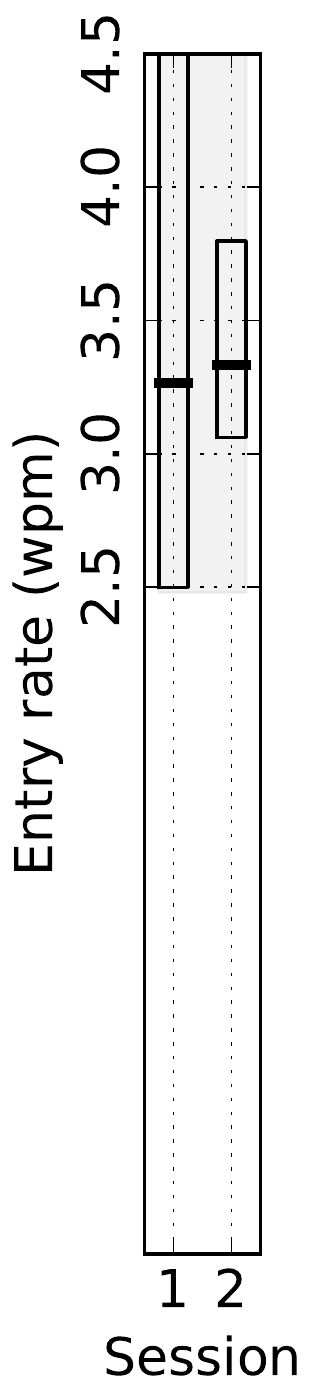}  \end{minipage} 
&
\begin{minipage}{1.35cm} \centering \includegraphics[height=5.6cm]{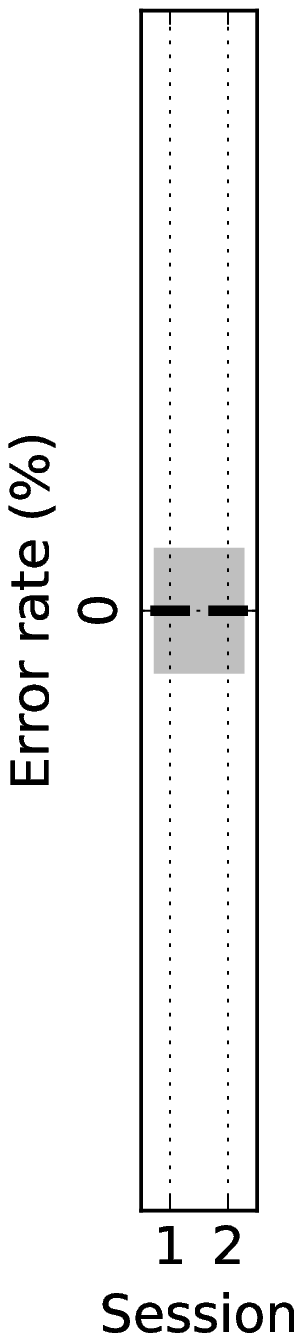}  \end{minipage} 
&
\begin{minipage}{1.35cm} \centering \includegraphics[height=5.6cm]{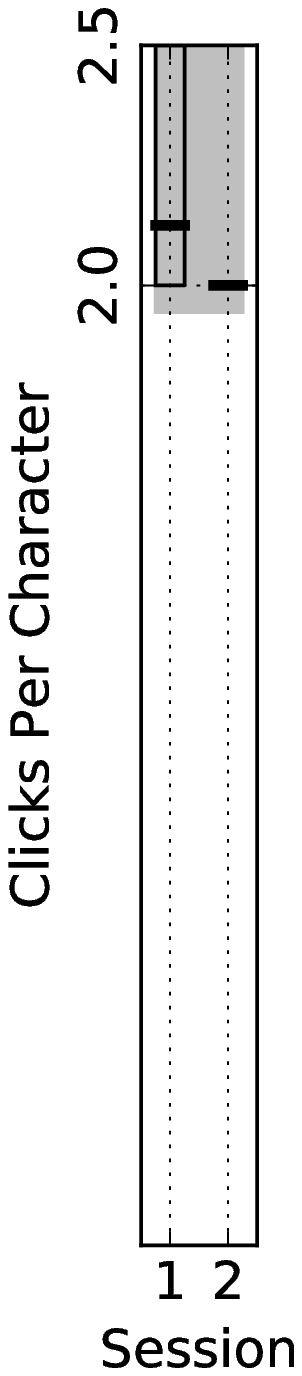}  \end{minipage} 
\end{tabular}
&
\begin{tabular}{c@{}c@{}c@{}}
\begin{minipage}{1.35cm} \centering \includegraphics[height=5.6cm]{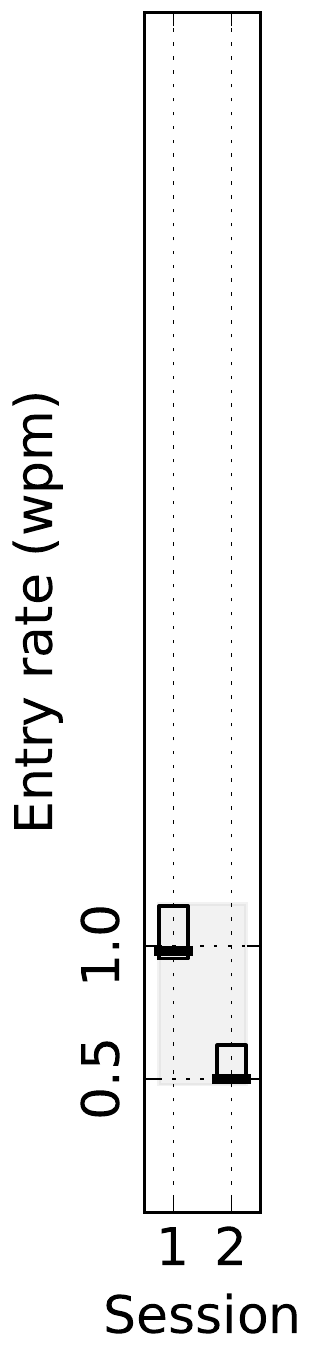}  \end{minipage} 
&
\begin{minipage}{1.35cm} \centering \includegraphics[height=5.6cm]{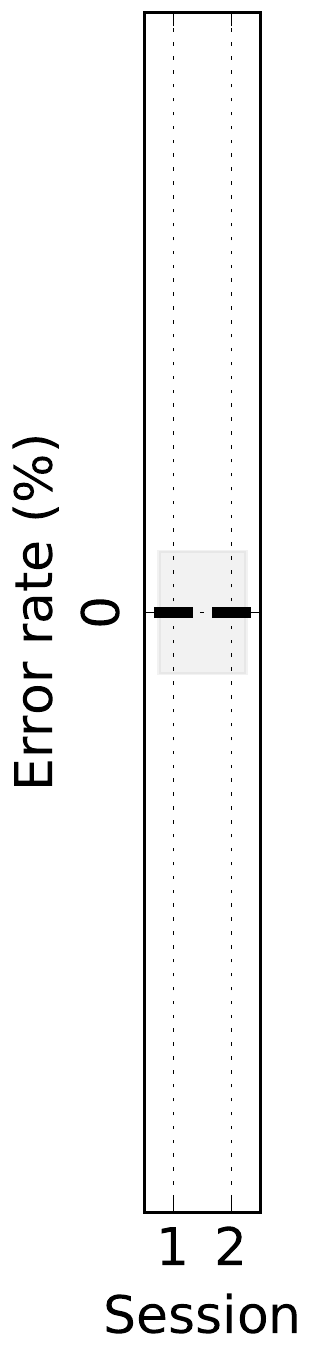} \end{minipage} 
&
\begin{minipage}{1.35cm} \centering \includegraphics[height=5.6cm]{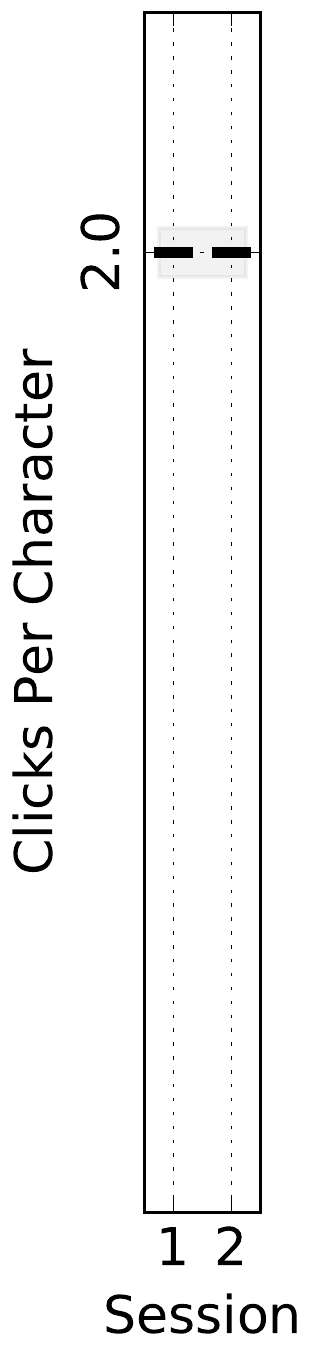}  \end{minipage} 
\\
\end{tabular}
\\(a) & (b) \\
\end{tabular}
\caption{ 
Box-and-whisker plots of the audio pilot study, indicating the 25th, 50th, and 75th percentiles.
Results for two sessions are shown in (a)-(b). Each session was 15 minutes long. Each session is numbered (x-axis).
(a) Results for a trained participant (with at least two hours of practise) communicating in an environment with little noise, where $T_{\mathrm{S}}$ was varied.  The first session was recorded when the user had at least one hour  of practise. The last session was recorded when the user could comfortably use the system blindfolded.
(b) Results for the same participant in (a) but simulating a non-speaking individual with motor  disabilities (by including synthetic noise and using the system blindfolded).   Session~1 presents a user   who can click precisely, but with some latency; 
\nu{$\Delta=0.8$}{s}, \nu{$\sigma=50$}{ms}, $\lambda=0$,  $f=0$,   and \nu{$T_{\mathrm{S}}=1.4$}{s}. 
The latency during Session~2 was  increased and some false positives were randomly generated with \nu{$\Delta=1.5$}{s}, \nu{$\sigma=50$}{ms}, $f=0.1$ and \nu{$\lambda=1/3$}{/s},  and \nu{$T_{\mathrm{S}}=2.1$}{s}. 
 Results from (b) can be compared to results from Session~2 in (a), as this was the result when the user had the most experience in a noiseless environment.
 }
\label{fig:audio_trials}
\end{figure}

At the beginning of the pilot study the participant represented an able-bodied novice user, i.e., the user could use the system error-free with \nu{$T_{\mathrm{S}}=500$}{ms}. Before the synthetic noise were added, the participant had enough training to represent and expert able-bodied user (with \nu{$T_{\mathrm{S}}=300$}{ms}).  The results of the pilot study are presented in Figure~\ref{fig:audio_trials}.
  
We focus  on simulations that can be used for a variety of facial gestures, where the user may have a long latency, but is able to click precisely (with small variance).
Motivated by our pilot study, the Grid2 reference manual~\cite{grid2} and measurements from existing literature (e.g.,~\cite{nomon}), the click-timing parameter ranges 
were set to  \nu{$\Delta \in [0.2, 3]$}{seconds}, and \nu{$\sigma \in [50, 200]$}{ms} during our simulations.

\subsection{Modelling Results}
\label{sec:simulation_results}

During all simulations we computed results for the pangram $W_\mathrm{x}=$``the quick brown fox jumps over the lazy dog .'', where  ``\_'' is used  to enter a space. This sentence is used to test algorithms in many papers, as it contains all the letters of the alphabet, and all the words frequently occur in English.  In all simulations $E=2, U=2, \kappa=5$.

\begin{figure*}[!!!!!htb]
\centering
\begin{tabular}{ccc}  
\begin{minipage}{4.5cm}\includegraphics[width=4.3cm]{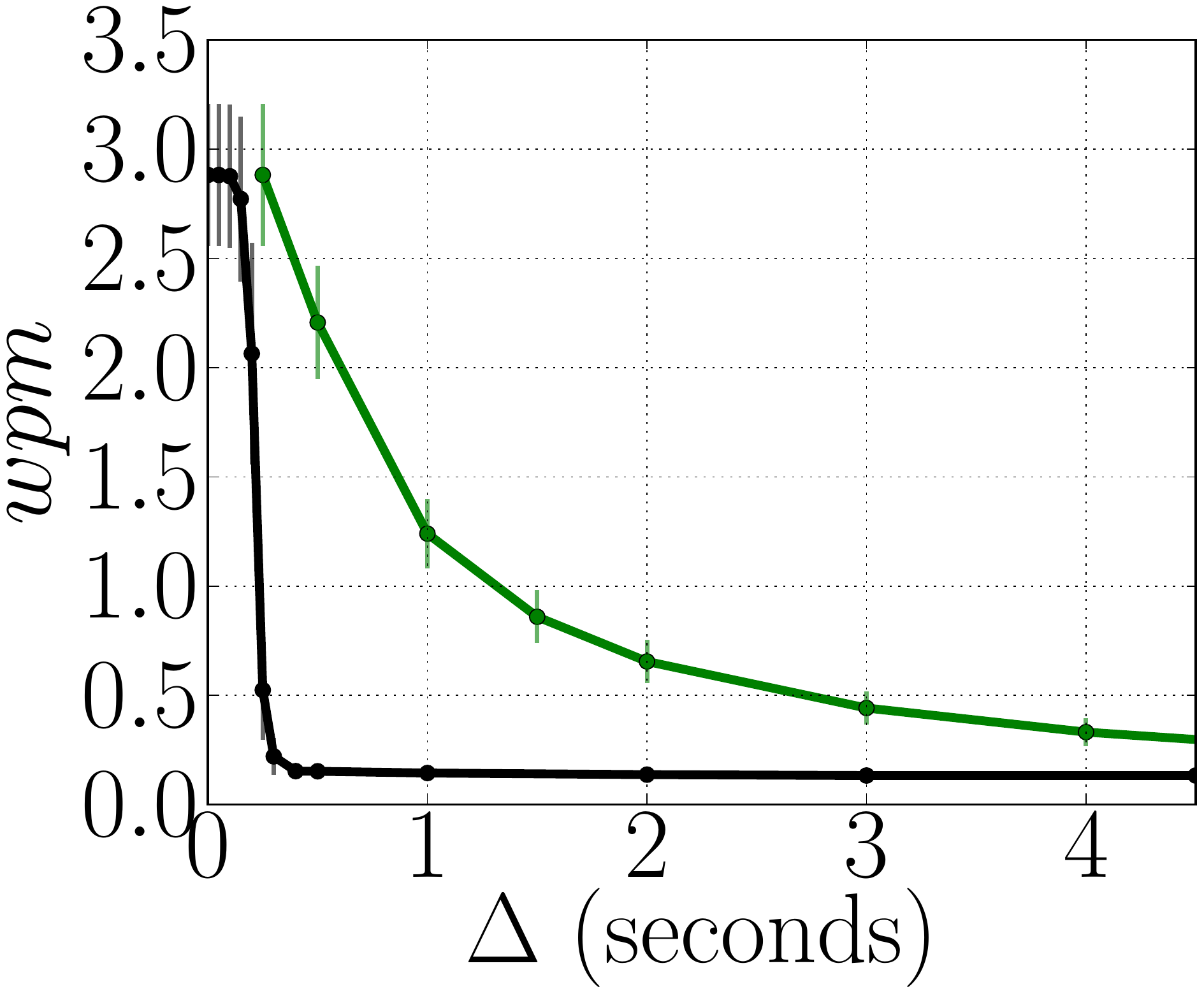} \end{minipage}
&
\begin{minipage}{4.5cm}\includegraphics[width=4.3cm]{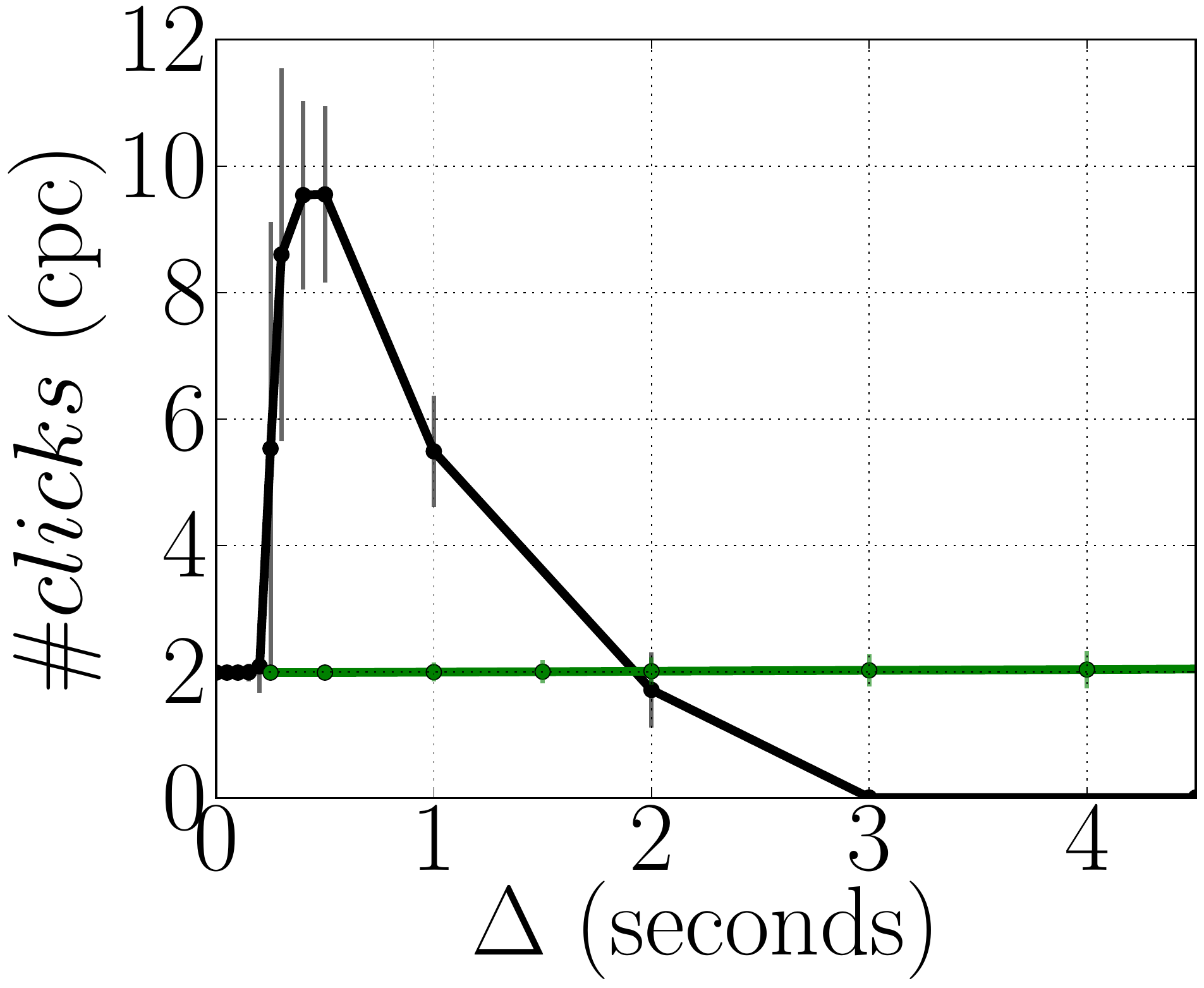} \end{minipage}
&
\begin{minipage}{4.cm}\includegraphics[width=4.3cm]{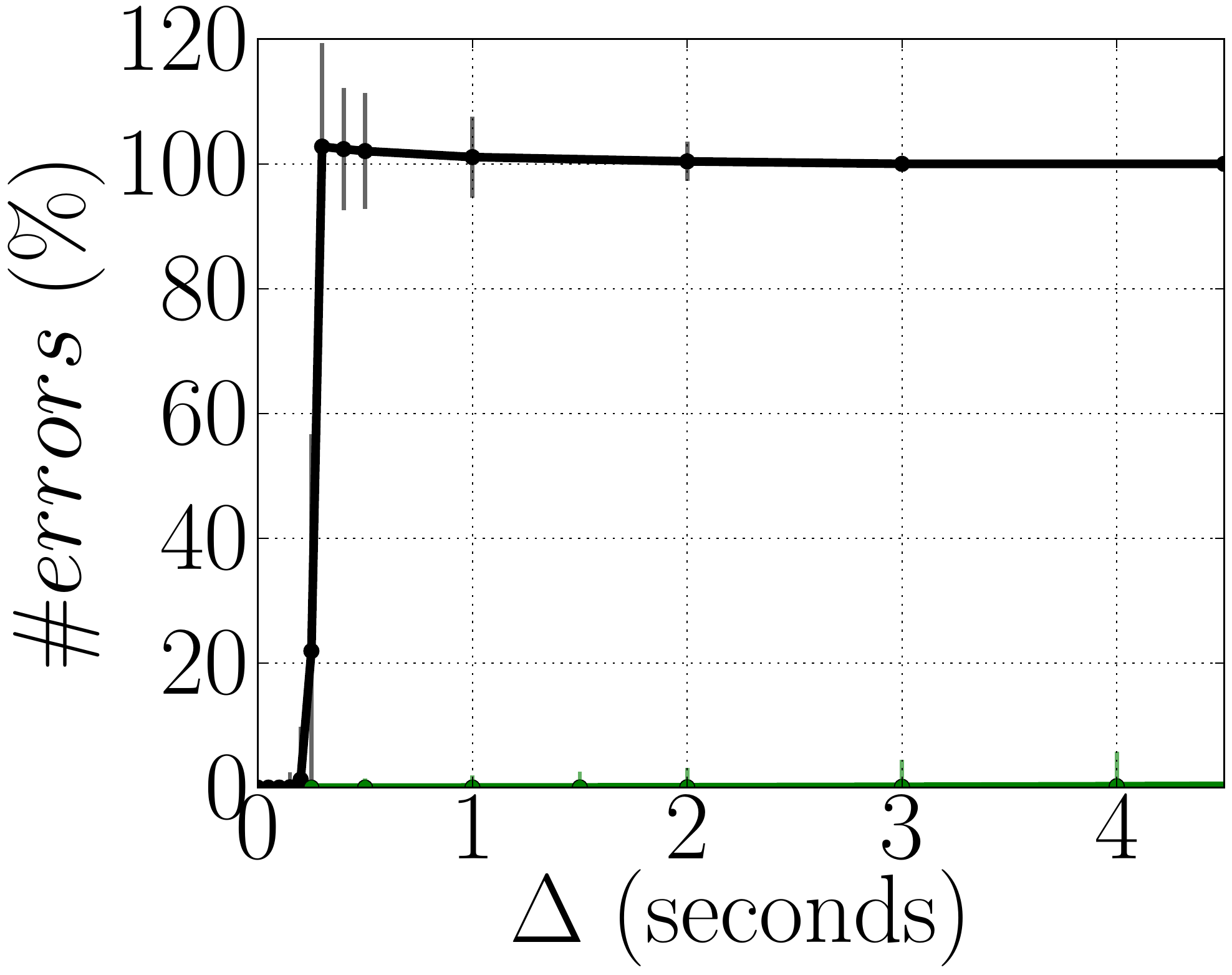} \end{minipage}
\end{tabular}
\caption{The click-timing delay ($\Delta$) is varied 
for \nu{$\sigma=50$}{ms}, $f=0.05$, \nu{$\lambda=0.001$}{/s} (on average, one false positive every 17 minutes).
 The black lines represent results for a fixed \nu{$T_{\mathrm{S}}=0.5$}{s} and variable $\Delta$. The green lines represent the results when 
 $T_{\mathrm{S}}$ is varied according to $\Delta$, where  \nu{$T_{\mathrm{S}} = \max(0.5,  \Delta + 3.0\sigma )$}{seconds} (i.e., avoiding error corrections due to a click-timing delay).    
  } 
\label{fig:click_time_delay}
\end{figure*}

The first simulation investigates the robustness to variations in latency~$\Delta$. Results are shown in 
Figure~\ref{fig:click_time_delay}. 

 The green graph indicates that it is significantly better to increase
$T_{\mathrm{S}}$ linearly with $\Delta$, instead of correcting the corresponding errors due to $\Delta$ which is 
too large compared to $T_{\mathrm{S}}$. This result is consistent with results from previous work in the literature~\cite{Koester2014}. 
 
 %The effect of considering the full click-timing distribution may become more pronounced if the scanning system is improved. One such improvement is to model the click-time distribution directly instead of increasing the scanning delay and accepting all clicks immediately. One can rather wait for all column/row scans to pass before making a decision on a cell selection. Depending on the user’s click-time latency, this can result in a significant boost in text-entry rate. 

 For validation purposes, note that \nu{$T_{\mathrm{S}}=1.4$}s, $wpm \approx 1$, and for \nu{$T_{\mathrm{S}}=2.1$}s, $wpm \approx 0.5$,
which is consistent with our pilot study text entry rates for the same $T_{\mathrm{S}}$, as shown in Figure~\ref{fig:audio_trials}(b).

\begin{figure}[!!!!!htb]
\centering
%\begin{tabular}{c@{}c@{}c@{}}  
%\begin{minipage}{2.8cm}\includegraphics[width=2.8cm]{pictures/simulations/trial5_0} \end{minipage}
%&
%\begin{minipage}{2.8cm}\includegraphics[width=2.8cm]{pictures/simulations/trial5_1} \end{minipage}
%&
%\begin{minipage}{2.8cm}\includegraphics[width=2.8cm]{pictures/simulations/trial5_2} \end{minipage}
%\\ 
%\multicolumn{3}{c}{ 
\begin{tabular}{l@{}l@{}}
\hspace{-0.5cm}
\begin{minipage}{3.8cm}\includegraphics[width=4.3cm]{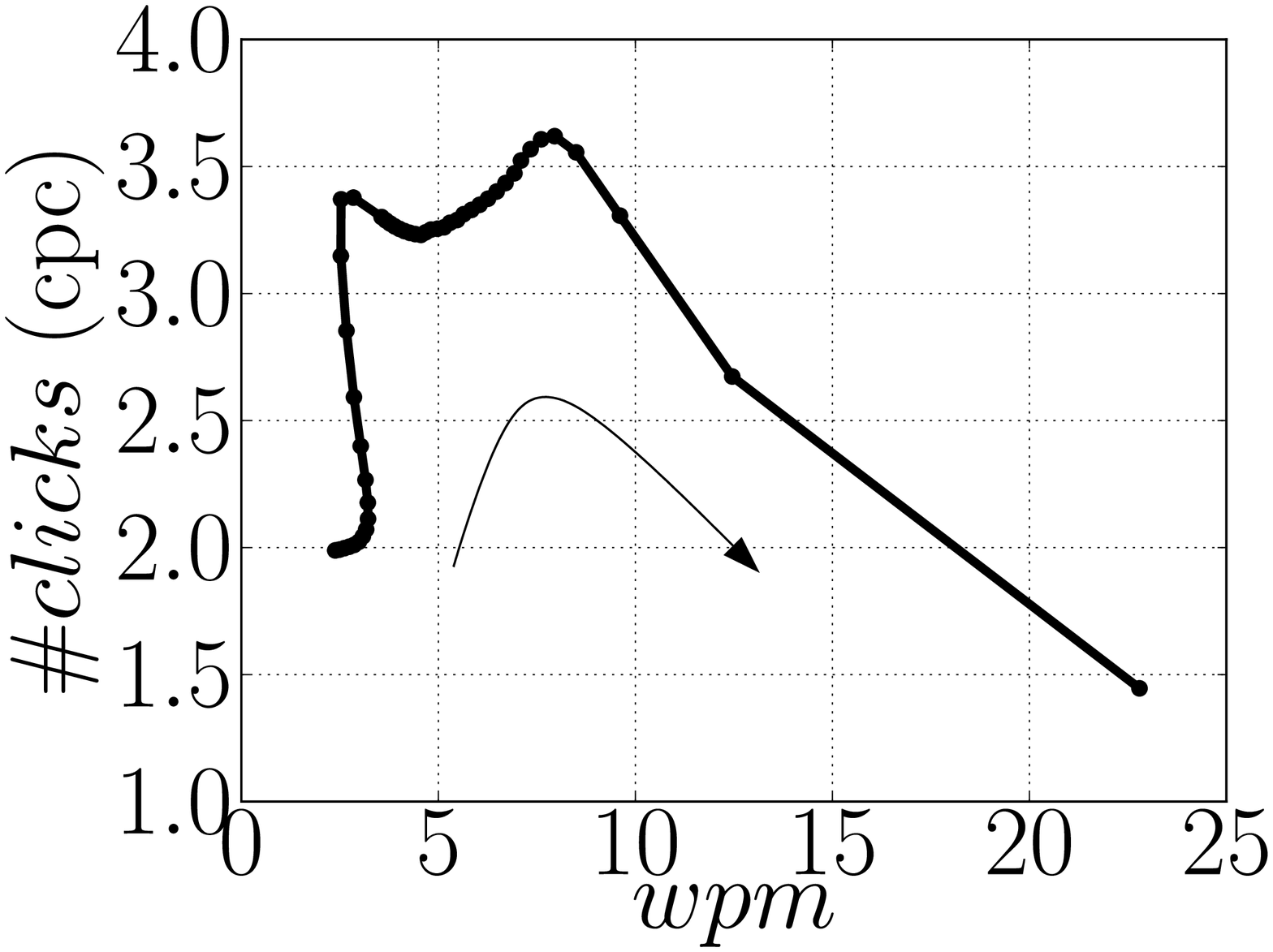} \end{minipage}
&
\hspace{0.5cm}
\begin{minipage}{4.3cm}\includegraphics[width=4.3cm]{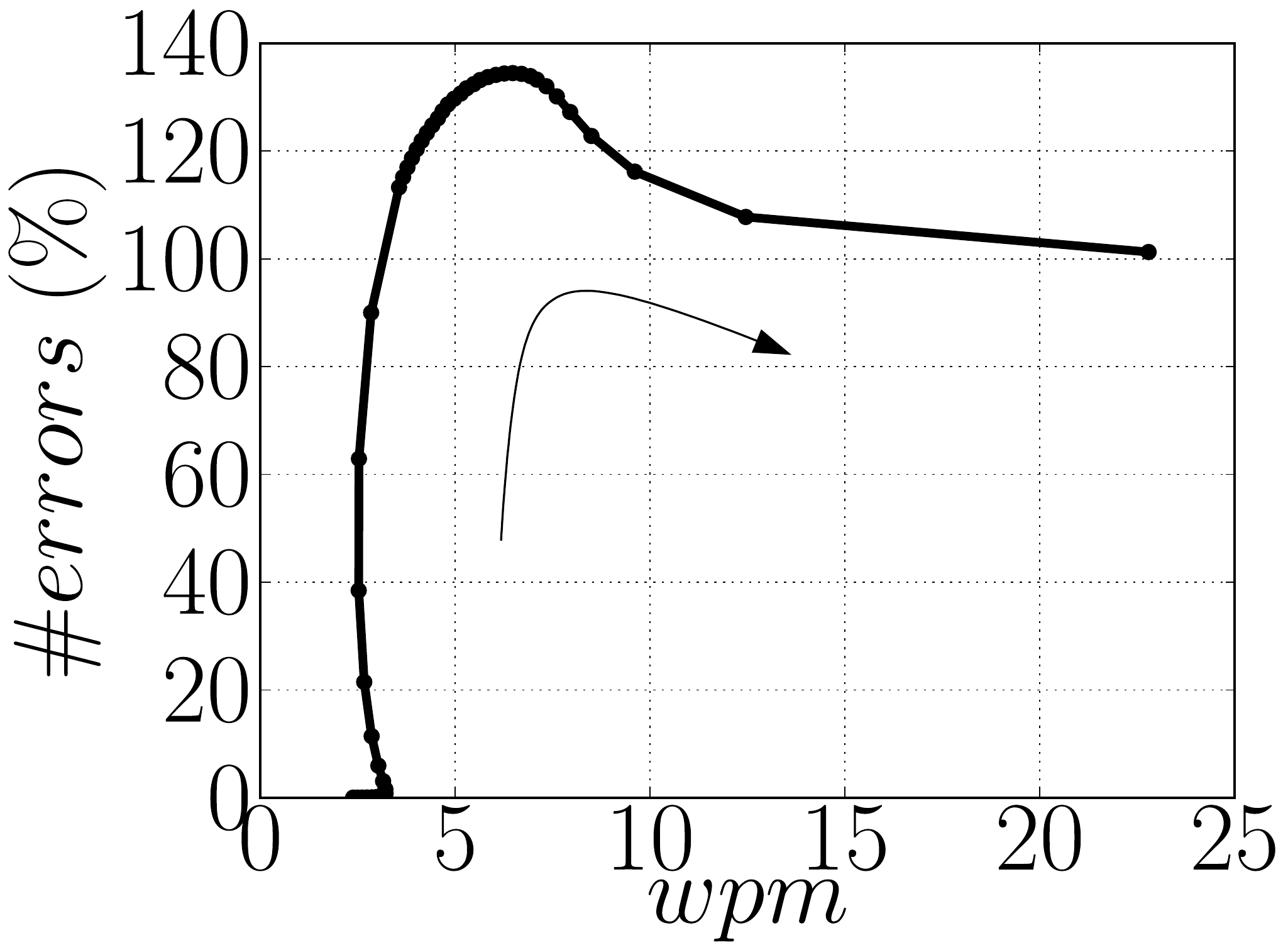} \end{minipage}
\\  
%\end{tabular}} 
\end{tabular}
\caption{The effect of varying the scanning delay $T_{\mathrm{S}}$ is investigated for \nu{$\Delta=0$}{s}, \nu{$\sigma=100$}{ms}, $f=0.05$, \nu{$\lambda=0.001$}{/s} (one false positive every 17 minutes).    Average output results  are shown in the direction of increasing $\Delta$ (starting at $\Delta=0$). The direction is indicated by the thin arrow line. } 
\label{fig:speed_increase}
\end{figure}

 The second simulation tested the influence of varying $T_{\mathrm{S}}/\sigma$.  $T_{\mathrm{S}}$ was set to a large value, and gradually decreased to a small value while keeping $\sigma$ fixed.
  Results are shown in
Figure~\ref{fig:speed_increase}.  

\begin{figure}[!!!!!htb]
\centering
%\begin{tabular}{c@{}c@{}c@{}}  
%\begin{minipage}{2.8cm}\includegraphics[width=2.8cm]{pictures/simulations/trial6_0} \end{minipage}
%&
%\begin{minipage}{2.8cm}\includegraphics[width=2.8cm]{pictures/simulations/trial6_1} \end{minipage}
%&
%\begin{minipage}{2.8cm}\includegraphics[width=2.8cm]{pictures/simulations/trial6_2} \end{minipage}
%\\ 
%\multicolumn{3}{c}{ 
\begin{tabular}{l@{}l@{}}
\hspace{-0.5cm}
\begin{minipage}{3.8cm}\includegraphics[width=4.3cm]{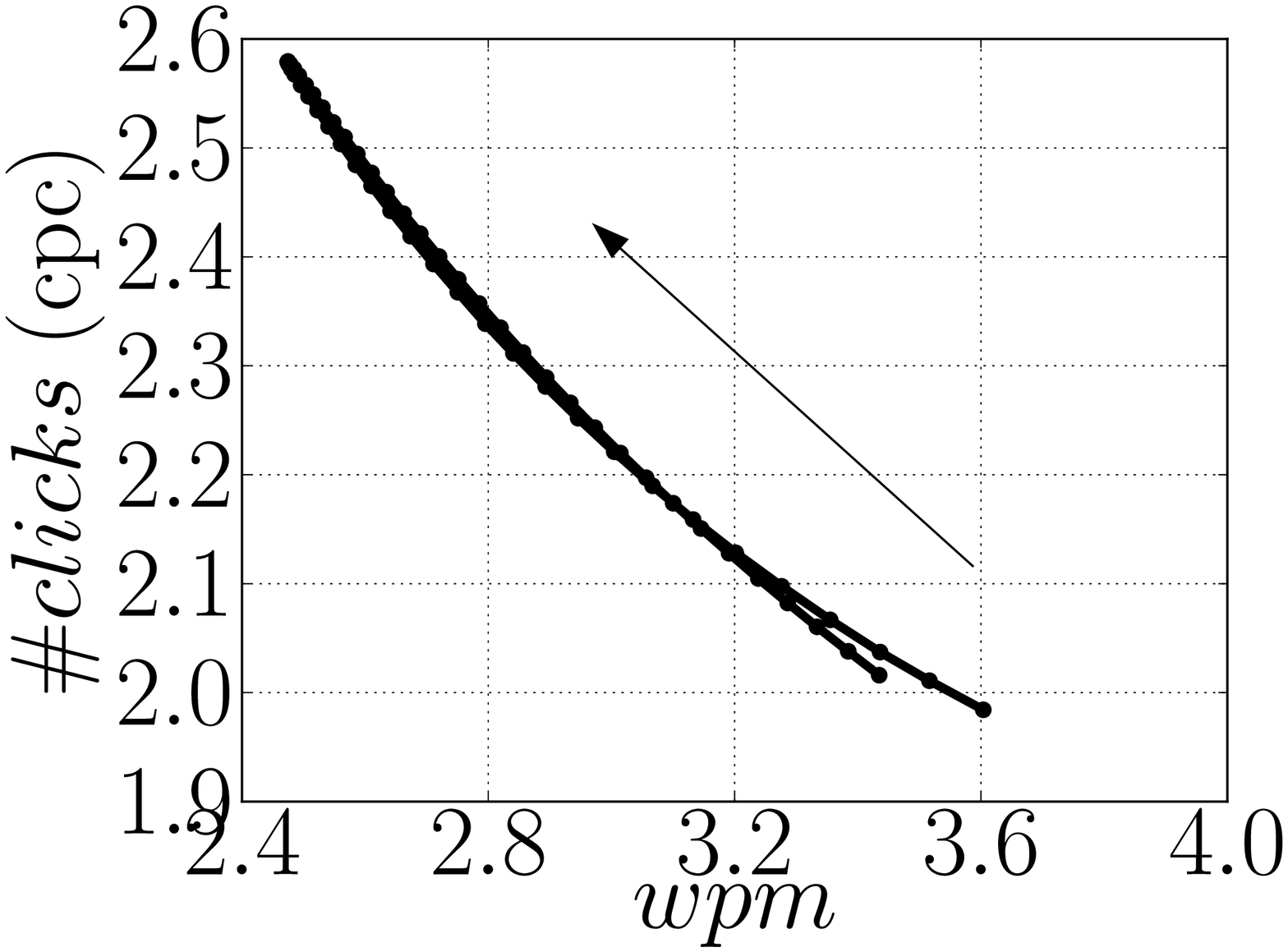} \end{minipage}
&
\hspace{0.5cm}
\begin{minipage}{4.3cm}\includegraphics[width=4.3cm]{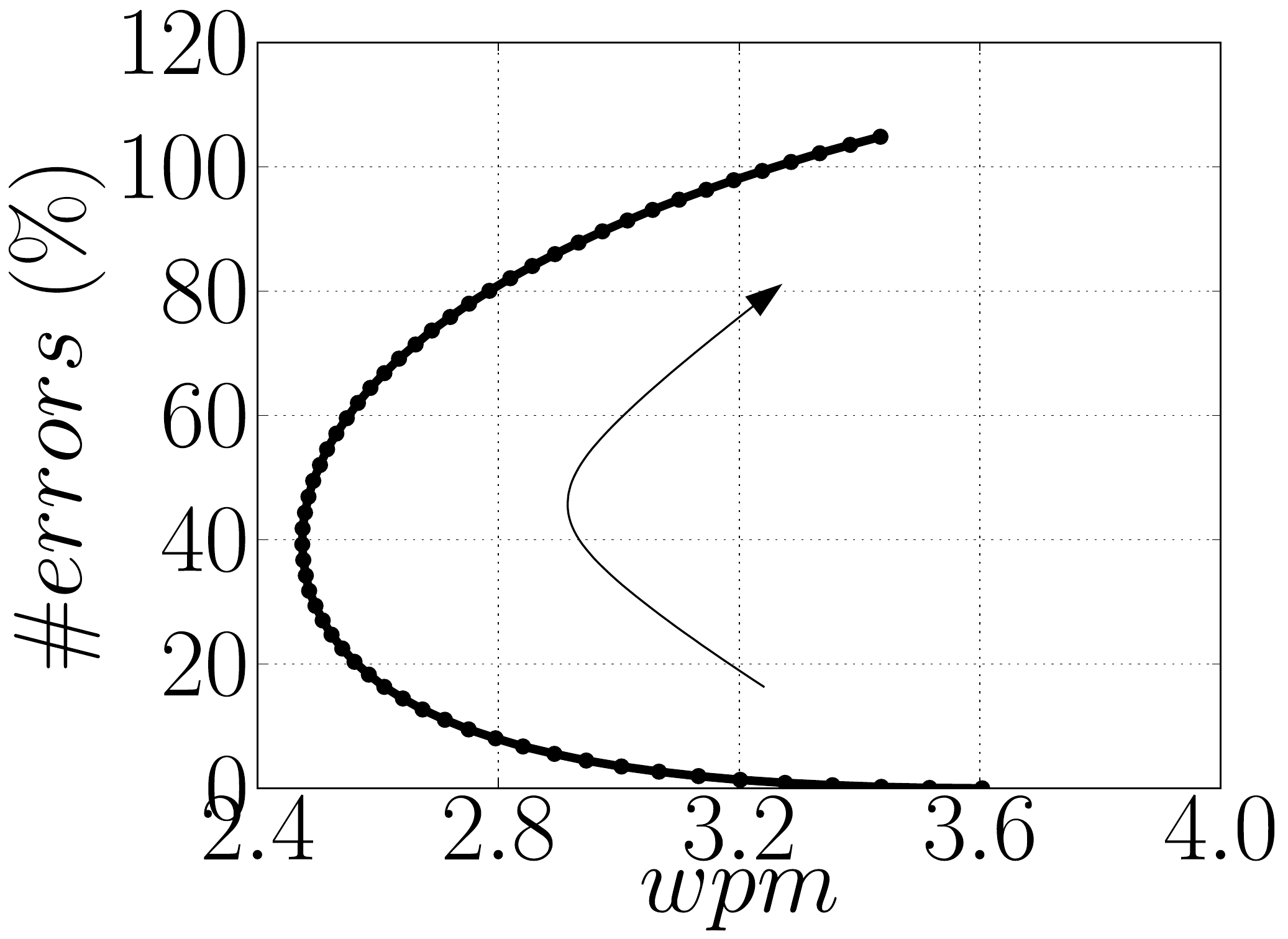} \end{minipage}
\\ 
%\end{tabular}} 
\end{tabular}
\caption{
The effect of varying $\lambda$ is investigated for \nu{$\Delta=400$}{ms}, \nu{$\sigma=50$}{ms}, $f=0.05$, and \nu{$T_{\mathrm{S}}=300$}{ms}.   Average output results  are shown in the direction of increasing $\lambda$ (starting at $\lambda=0$). The direction is indicated by the thin  arrow line.
   } 
\label{fig:false_positives}
\end{figure}

  Figure~\ref{fig:speed_increase} indicates that many erroneous characters eventually lead to the system failure, as the text entry rate and error rate are both high. There is a small working range for the program, where a reasonable accuracy ($\#error < 5\%$)  and click rate  ($\#clicks \leq 2$)  
can be expected. The latter performance range was measured for $T_{\mathrm{S}} / \sigma \geq 3$. A practical example would be \nu{$T_{\mathrm{S}}=300$}{ms} and \nu{$\sigma \approx 100$}{ms} when used by an expert able-bodied user. This is easily achievable with our (and other) typical scanning software, especially when other software is used in visual mode. When 
 \nu{$T_{\mathrm{S}}=500$}{ms} (used by a novice able-bodied user), \nu{$\sigma < 166$}{ms} is required, and when \nu{$T_{\mathrm{S}} = 1$}{s} (used by an expert non-speaking individual with motor disabilities),  \nu{$\sigma < 333$}{ms} is required. It was noted that the standard deviation increases as a user experiences fatigue (when using a facial gesture such as blinking to communicate). This highlights the potential benefits of taking the full click-time distribution into account to determine $T_{\mathrm{S}}$ dynamically.
  
The third simulation tested the effect of false positives. Results are shown in
 \figref{fig:false_positives}. The point where the scanning system started to degrade substantially was measured at \nu{$\lambda > 0.05$}{/s}. It is unlikely that more error correction actions will help to increase robustness to a high false positive rate. If the scanning system is able take the full noise distributions into account, one might be able to ignore certain clicks avoiding to undo them afterwards.

\section{Improving Scanning using a Noise Model}
\label{sec:improving_grid}

In this section we present a novel modification to a standard scanning system to potentially benefit users with long response times.
We call the proposed method the {\em fast-scan} method, and the standard scanning method the {\em slow-scan} method. The standard scanning system 
is based on the most basic system the standard scanning system Grid2 recommends, as mentioned before. 

We make use of our model for scanning systems to measure the efficacy of making use of an explicit noise model (in the form of a probability distribution) as part of the text entry method interface. For illustrative purposes, false positives are ignored, as computing  their probabilities for the fast-scan method involves a bit more work, and is the topic of a future study.

Assuming, $\Delta$, the average click-timing delay is known (it can be estimated by measurement), the idea is this: instead of immediately generating a switch event when a click is received, 
defer until the end of a group scan to make a decision. Then use Bayesian inference to infer the user's intentions. As a start, choose the most probable cell according the well-known maximum a posteriori decision rule: 
\begin{equation}
v^{*} =  \argmax\limits_{v}  P ( t - \Delta  \mid \boldsymbol{\theta}, v,v'   )
\end{equation}
which is the result when maximising over the posterior probability $ P( v \mid \boldsymbol{\theta}, t  - \Delta , v'  )$,  and assuming a uniform prior over $v$, i.e, $P( v) = 1/V$. Waiting until the end of a group scan enables one to subtract the average response time $t - \Delta$ before doing inference.

From a simulation point of view, the states in our state diagram are the same as for the slow-scan method, but the transition probabilities are different. More specifically,
 $p_{0}(v', v) = 1.0$ if $v' < V$. That is, the miss probability is one for all cells except for the last one in the group.
At $v'=V$, i.e., at the end of the last scan in a group,  $p_{1}(v', v)$ from \eqref{eq:p1} is computed for all cells, as before but with~$T_{\mathrm{S}} = T_{v'}$,
where $T_{v'}$ is the scanning delay associated with cell~$v'$. However, after computing  $p_{1}(v', v)$ as before,  compute the final $p_{0}(v', v) = 1 -  \sum_{v'=1}^{V} p_{1}(v', v)  $.

As before, $q(v',v)$  represents the probability of error associated with the click-timing distribution (the probability to click while scanning cell~$v'$, with the intention to select cell~$v$). 
To compute  $q(v',v)$
use   \eqref{eq:overlap}, and set $T_{\mathrm{S}} = T_{v'}$.  

$ P ( t  - \Delta \mid \boldsymbol{\theta}, v,v'  )$, which is used to specify $q(v',v)$, can be any distribution that is continuous in time, but for comparative purposes it is initially assumed to be Gaussian.  When using a Gaussian, one measures the difference in performance if both methods know that the received click times will fall within a specified {\em range}.

Since false positives are ignored, the only other addition to the algorithm that is necessary, is that $r_{n'} \in \{1,2, k_{\Delta}\}$  (see \eqref{eq:scan_prob_state_probs}), where $k_{\Delta}= \mathrm{ceil}( T_\mathrm{S} / T_\mathrm{S}^{\mathrm{FAST}})$.

\figref{fig:grid2_wait_state_diagram} depicts the reduced set of transition links (compared to \figref{fig:grid2_wait}).

The simulation of the fast scan method requires an extra parameter:
\begin{equation}\label{eq:theta_fast}
{\gp}^{\mathrm{FAST}} = \{ \gp,  T_\mathrm{S}^{\mathrm{FAST}}  \},
\end{equation} 
%\Delta, \sigma, f, \lambda, T_\mathrm{S}, T_\mathrm{S}^{\mathrm{FAST}} U, E, \kappa \}.
where $T_\mathrm{S}^{\mathrm{FAST}}$ is the scanning delay of all the cells except the last cells in the group. If an additional ``tick" sound is included when using the system in audio mode, the scanning delay of the first cell will be  \nu{$2T_\mathrm{S}^{\mathrm{FAST}}$}{seconds}.  The scanning delay of the last cell in a group is  \nu{$T_\mathrm{S}$}{seconds}. In general, we let  $T_\mathrm{S} = \Delta- 3\sigma $, which should ensure a small probability of error, where $T_\mathrm{S}$  is the scanning delay of the slow-scan method. To summarize:
\begin{align} \label{eq:scanning_delays}
T_{v'}  &= \begin{cases}  
       2T_\mathrm{S}^{\mathrm{FAST}} &  \text{if } v'=0,   \nonumber  \\
       T_\mathrm{S}^{\mathrm{FAST}}  &  \text{if } v' < V,  \nonumber \\
        T_\mathrm{S}                 &  \text{if } v'=V.
 \end{cases} 
\end{align}

The goal is to get an indication of the performance in the case where  both the old and new interface know exactly what the noise distribution is.

\subsection{Modelling the Average Response Time}

 In the first set of simulations, the effect of false positives and the standard deviation of the click-timing response is ignored for illustrative purposes. 
As a first example, assume that the false negative probability $f=0$, the false positive rate $\lambda=0$, and that $\sigma \approx 0$, so that $\mathcal{N}(t - \Delta ; \mu_v  , \sigma) \approx \delta(t -\Delta -  \mu_v )$. In this case, the number of scans can be counted  manually  for both methods and compared to the results from our simulation model for verification purposes. The expected text entry rate (number of scans) from the simulation should converge to the same value with a probability close to 1.0. 

By deferring the decision to the end of a group scan and trusting the noise model, the relationship between the scanning- and click-timing delay is decoupled at all cells except for the last one in the group. The last scanning delay in the group will have to be at least as long as the average click-timing delay (when ignoring the false- positives and the standard deviation of the click-timing).

%\begin{landscape}
\begin{figure*}[!!!!!t]
\begin{tabular}{l@{}}   

{
\setlength{\extrarowheight}{-1mm}
\input{figures/grid2.tikz}
}

\\ (a) \\

\begin{tabular}{l@{}l@{}l@{}l@{}}   
\begin{minipage}{4.3cm} \vspace{1mm} \includegraphics[width=4.3cm]{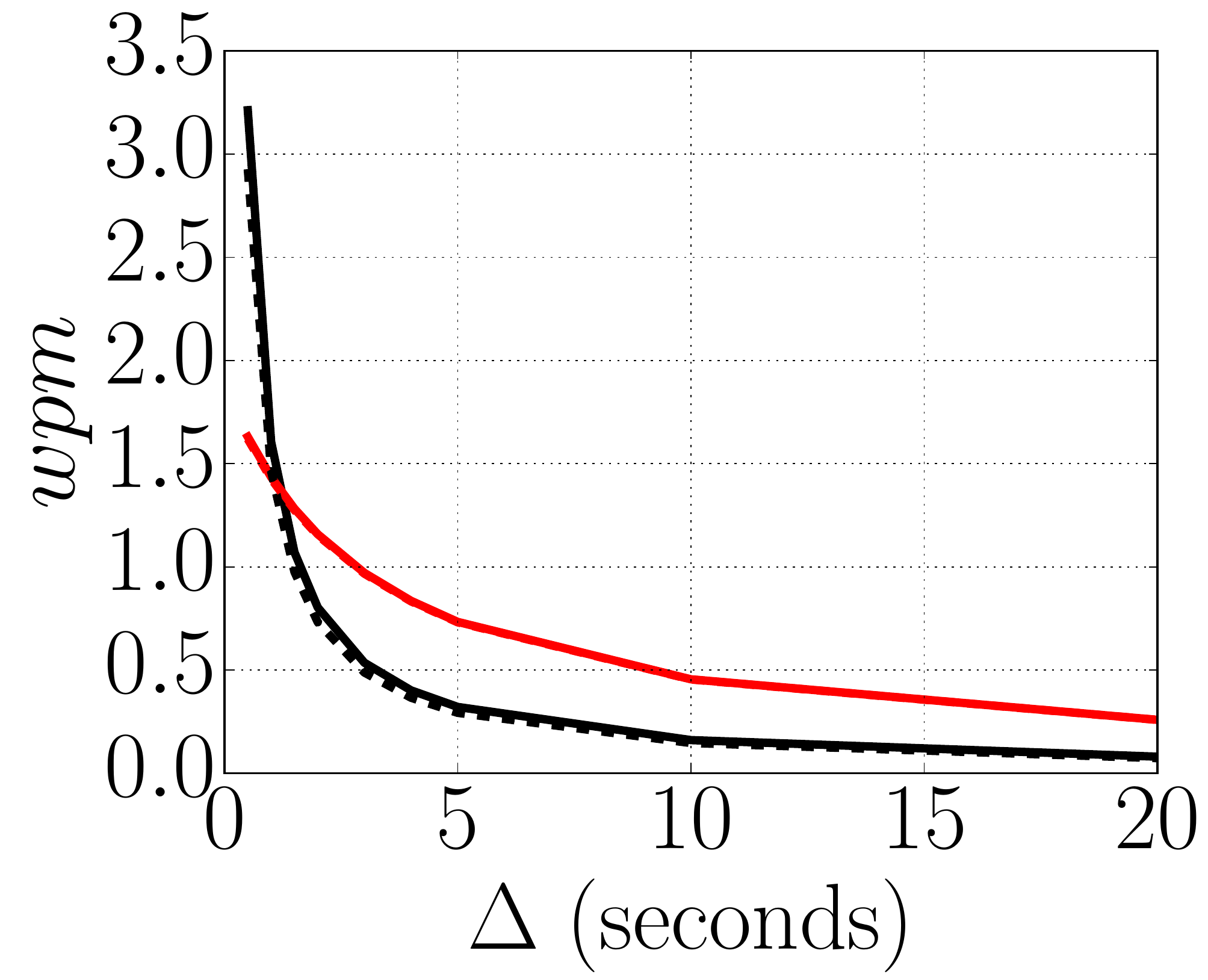} \end{minipage}
& 
\begin{minipage}{4.3cm} \vspace{1mm} \includegraphics[width=4.3cm]{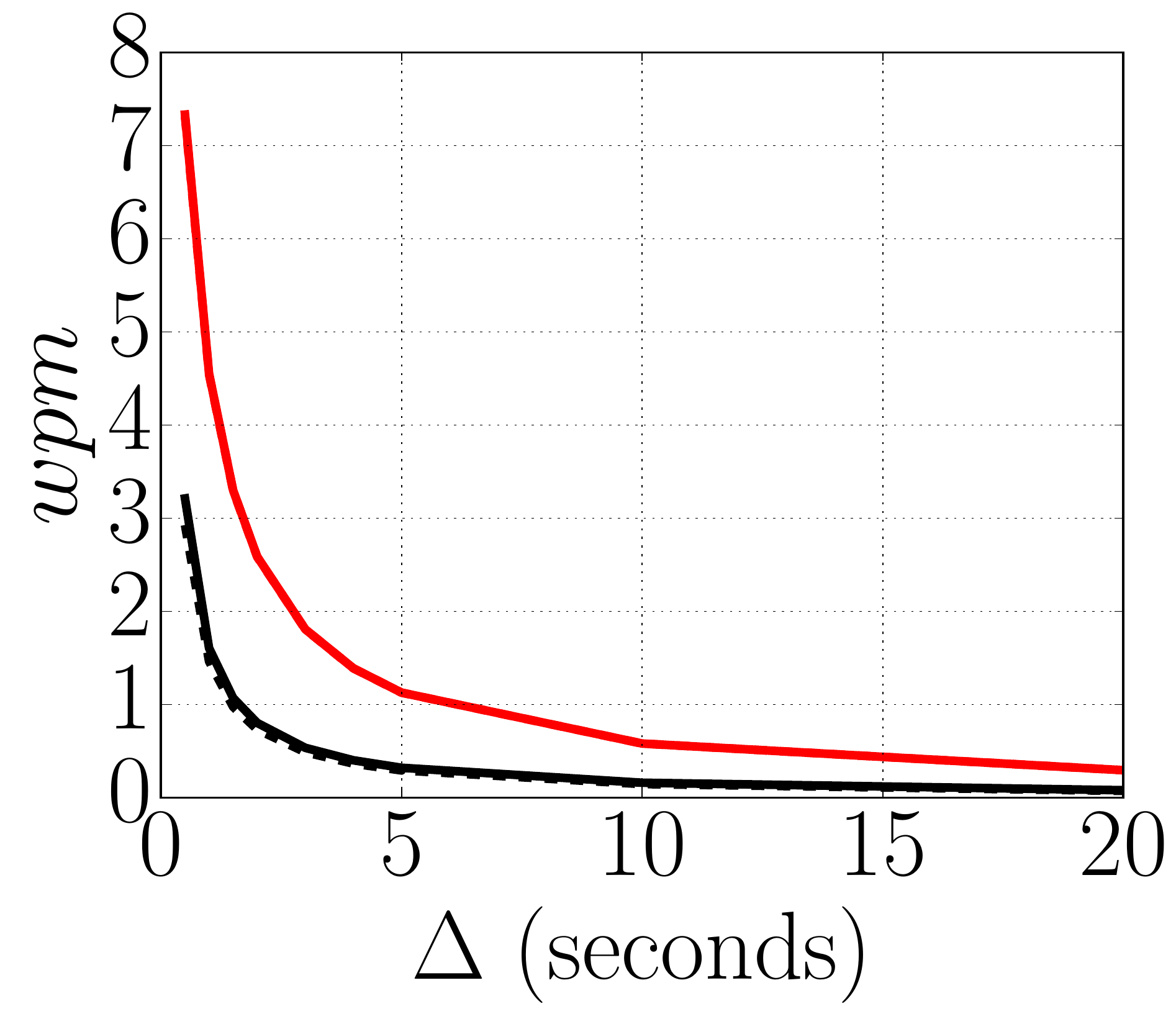} \end{minipage}
&
\begin{minipage}{4.3cm} \vspace{1mm} \includegraphics[width=4.3cm]{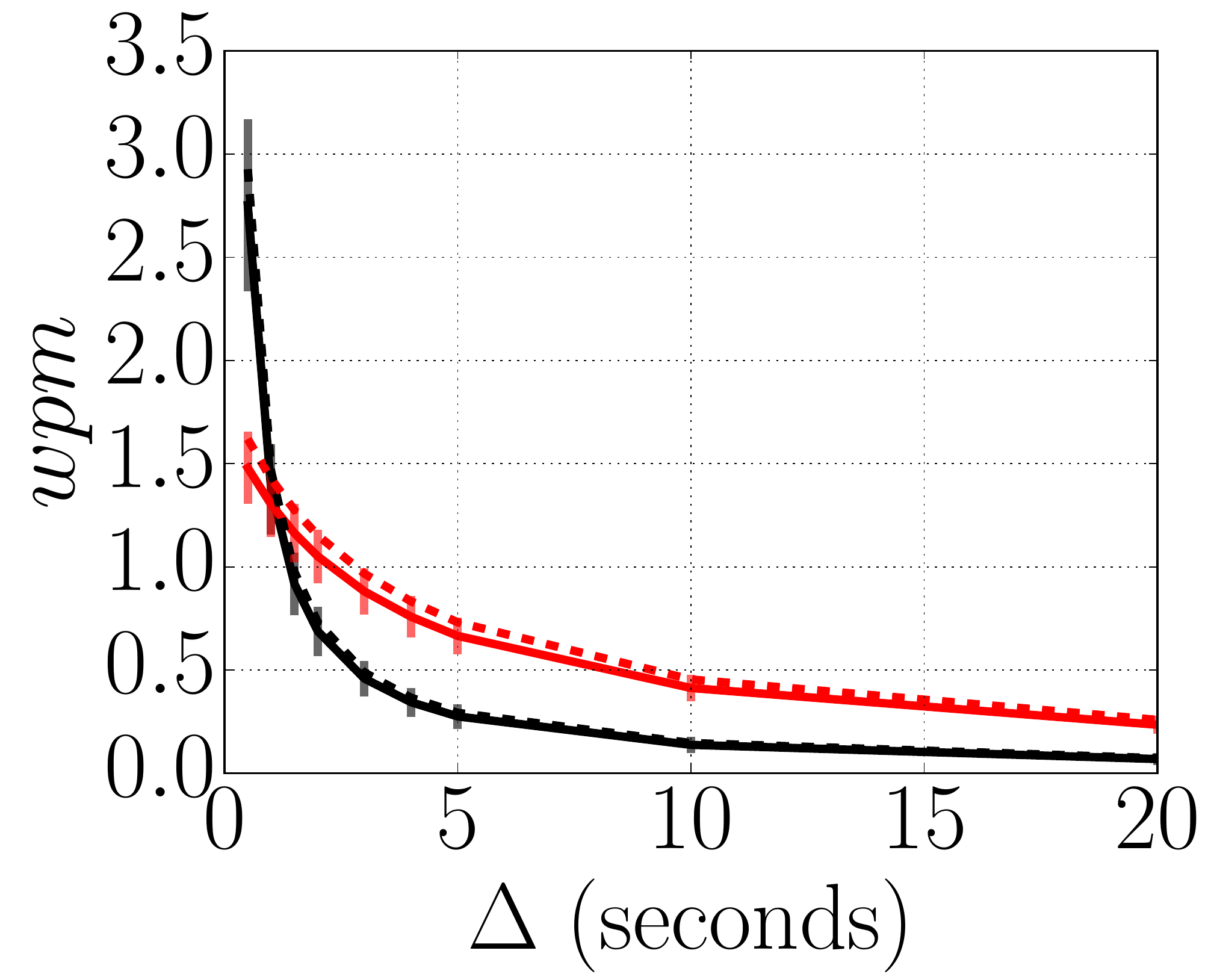} \end{minipage}
&
\begin{minipage}{4.3cm} \vspace{1mm} \includegraphics[width=4.3cm]{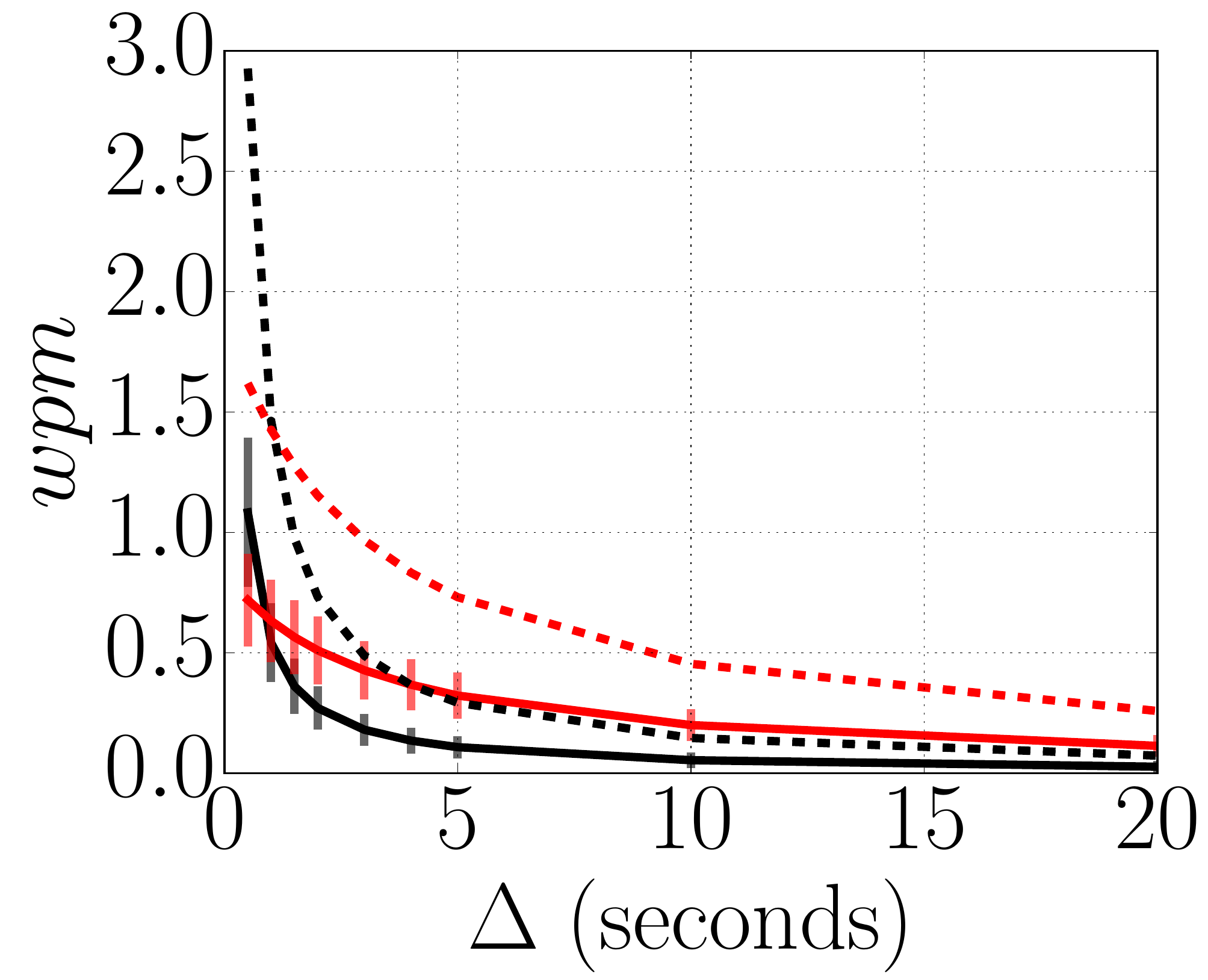} \end{minipage}
\\
\end{tabular}
\\
\begin{tabular}{cccc} 
\centering
\begin{minipage}{4.3cm}{(b)} \end{minipage} & 
\begin{minipage}{4.3cm}{(c)} \end{minipage} & 
\begin{minipage}{4.3cm}{(d)} \end{minipage} &
\begin{minipage}{4.3cm}{(e)} \end{minipage}  \\

\end{tabular}
\end{tabular}
\caption{(a) The state diagram for the fast-scan method.  (b)-(e) Comparing the slow-scan (black) and fast-scan (red) methods. In all cases $\sigma \approx 0, \lambda=0, U=2, E=2, \kappa=10$, and $\Delta$ is systematically increased to visualise the effect on the text-entry rate, and $T_\mathrm{S} = \Delta$. The dashed lines indicate the maximum text-entry rate in noise-less conditions. (b) Comparing the two methods in noiseless conditions with \nu{$T_\mathrm{S}^{\mathrm{FAST}}\approx 500$}{ms}, (c) \nu{$T_\mathrm{S}^{\mathrm{FAST}}\approx 50$}{ms},  (d) \nu{$T_\mathrm{S}^{\mathrm{FAST}}\approx 500$}{ms},  $f=0.1$, (e) \nu{$T_\mathrm{S}^{\mathrm{FAST}}\approx 500$}{ms},  $f=0.5$.  } 
\label{fig:grid2_wait_state_diagram}
\end{figure*}

The simulation is designed so that the number of errors and received clicks are the same for both methods (i.e., two clicks will be required per character, and the probability of error is negligible). We therefore isolate only the text entry speed (measured in exactly the same noise conditions). As an example: for the two-cell configuration associated with \figref{fig:grid2_wait} one can expect nine scans to write ``a\_" (including one extra scan per group for the ``tick" sound) when using the slow-scan method. Using the fast-scan method, one can expect twelve scans, where four of these will be slow ($T_\mathrm{S}$ seconds) and eight will be fast  ($T_\mathrm{S}^{\mathrm{FAST}}$ seconds).
 
The result for the more comprehensive example from \secref{sec:simulation_results} is shown \figref{fig:grid2_wait_state_diagram}(b)-(c) i.e., $W_\mathrm{x}=$``the quick brown fox jumps over the lazy dog .'', and using the layout from \figref{fig:grid2}. The difference between the simulations (solid lines) and the ground-truth results (dashed lines) is negligibly small. In \figref{fig:grid2_wait_state_diagram}(b) \nu{$T_\mathrm{S}^{\mathrm{FAST}}\approx 500$}{ms}. The simulation starts with \nu{$\Delta=500$}{ms} and \nu{$T_\mathrm{S}\approx 500$}{ms}. In the beginning the slow-scan method is faster than the fast-scan method. However, the fast-scan method quickly surpasses the slow-scan method. 

Increasing    $T_\mathrm{S}$  affects the fast-scanning method, but to a lesser degree. The text entry speed does however, also decrease  exponentially. \figref{fig:grid2_wait_state_diagram}(c) shows what happens when the \nu{$T_\mathrm{S}^{\mathrm{FAST}} = 50$}{ms}, while keeping \nu{$T_\mathrm{S}\approx 500$}{ms} with exactly the same experimental setup. Again the ground-truth and test results are the same. It is astounding how much faster the fast-scanning method is, theoretically allowing a person with a \nu{500}{ms} delay to write at more than \nu{7}{wpm}.

\figref{fig:grid2_wait_state_diagram}(d) depicts the effect of including a false negative rate of 10\%, ie., $f = 0.1$. We expect the text entry rate to reduce in both methods, but that the effect should be similar for both methods. Both results are indeed close to their baseline values and the error bars (one standard deviation) are similar for both (length of the small vertical lines). 
Like before, the error bars provide a range of possible outcomes that one can expect. If the expected text-entry speed is the same as the baseline, but the error bars are large, it is an indication that the performance is significantly worse than the performance of the baseline. \figref{fig:grid2_wait_state_diagram}(e) illustrates this, where $f = 0.5$: the false negative rate is significantly higher compared to \figref{fig:grid2_wait_state_diagram}(d) resulting in larger error bars and a worse average performance compared to the baseline.

%Now that we can decouple the user's average response time from the scanning delay, it will also take some creative thinking to determine how to represent the alphabet very quickly to the user. Can we present a letter to the user at \nu{50}{ms}, so that it is still audible? A click sound is typically \nu{12}{ms}, but it is difficult to record a letter at under  \nu{200}{ms}. When using an audio interface, all letters are usually represented serially to the user. Can the stimulus be parallelised in some way, similar to a visual interface? This is also an issue that we are currently investigating. 

\subsection{Modelling the Standard Deviation of the Response Time and Extensions to other Click-Timing Distributions}
 
We have previously shown that the fast-scan method effectively translates the click-timing distribution by subtracting the measured click-timing delay~$\Delta$. However, the model cannot capture any knowledge of the standard deviation. If increasing the scanning delay to reduce to probability of error has to be avoided, it will be necessary to identify another way to scale the effect of the  standard deviation. To achieve this the model requires more degrees of freedom. This can e.g., be achieved by deferring the decision to the end of a letter, i.e., after two group scans. Hence the user has two chances to click per letter. Two click-timing distributions must be considered, and their means and standard deviations must be consistent.
 The final click-timing distribution will result from a convolution of the two individual ones.
%If thought is given to the layout one can end up with a much narrower final distribution that makes a decision regarding the user's intention, compared to the result when considering only one distribution. In this way one could also increase the resilience to 
%false positives. This  topic is currently being researched.

\begin{figure}[!!!!!t]
\begin{tabular}{c@{}c@{}}   
\begin{minipage}{4.3cm} \vspace{1mm} \includegraphics[width=4.3cm]{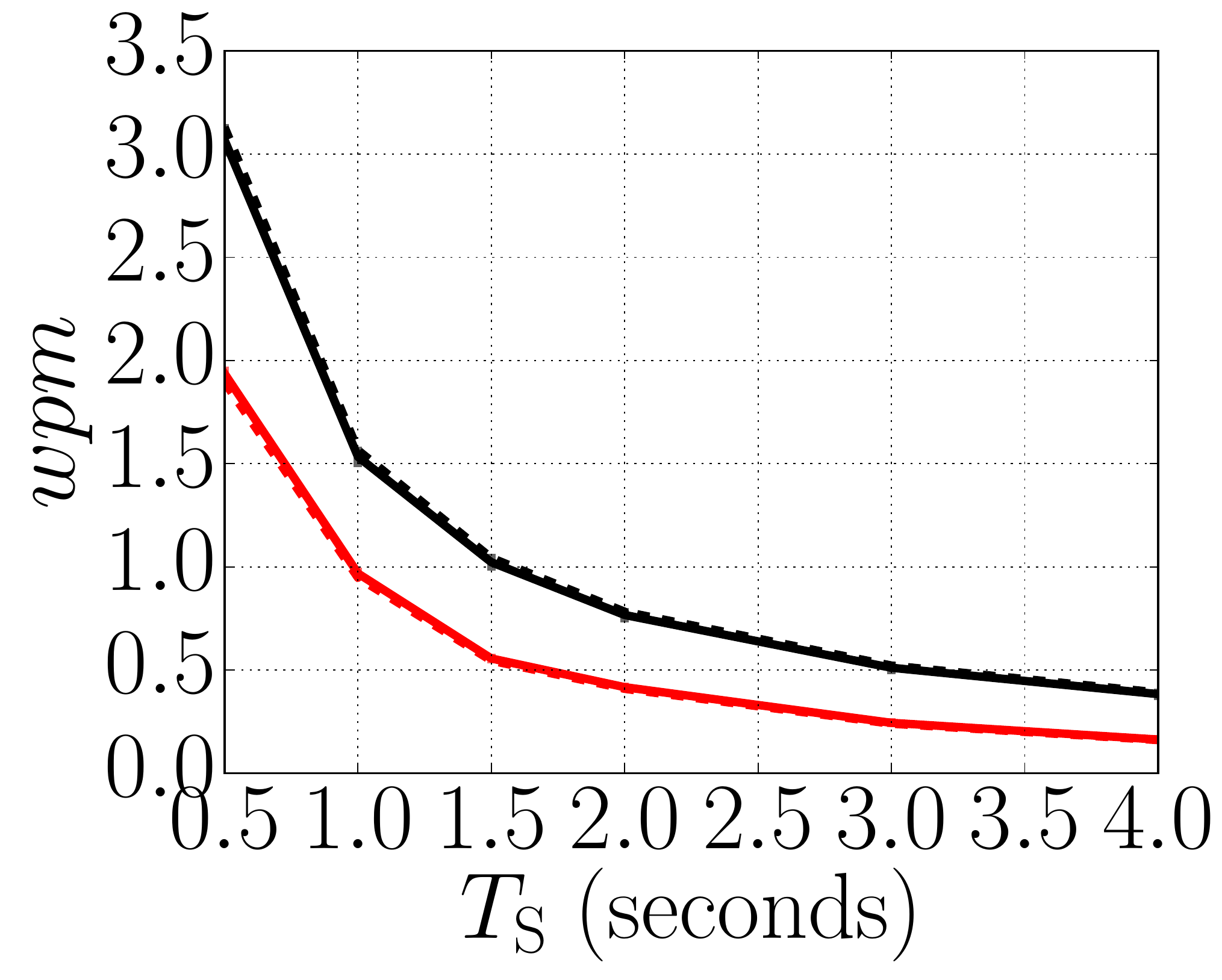} \end{minipage}
& 
\begin{minipage}{4.3cm} \vspace{1mm} \includegraphics[width=4.3cm]{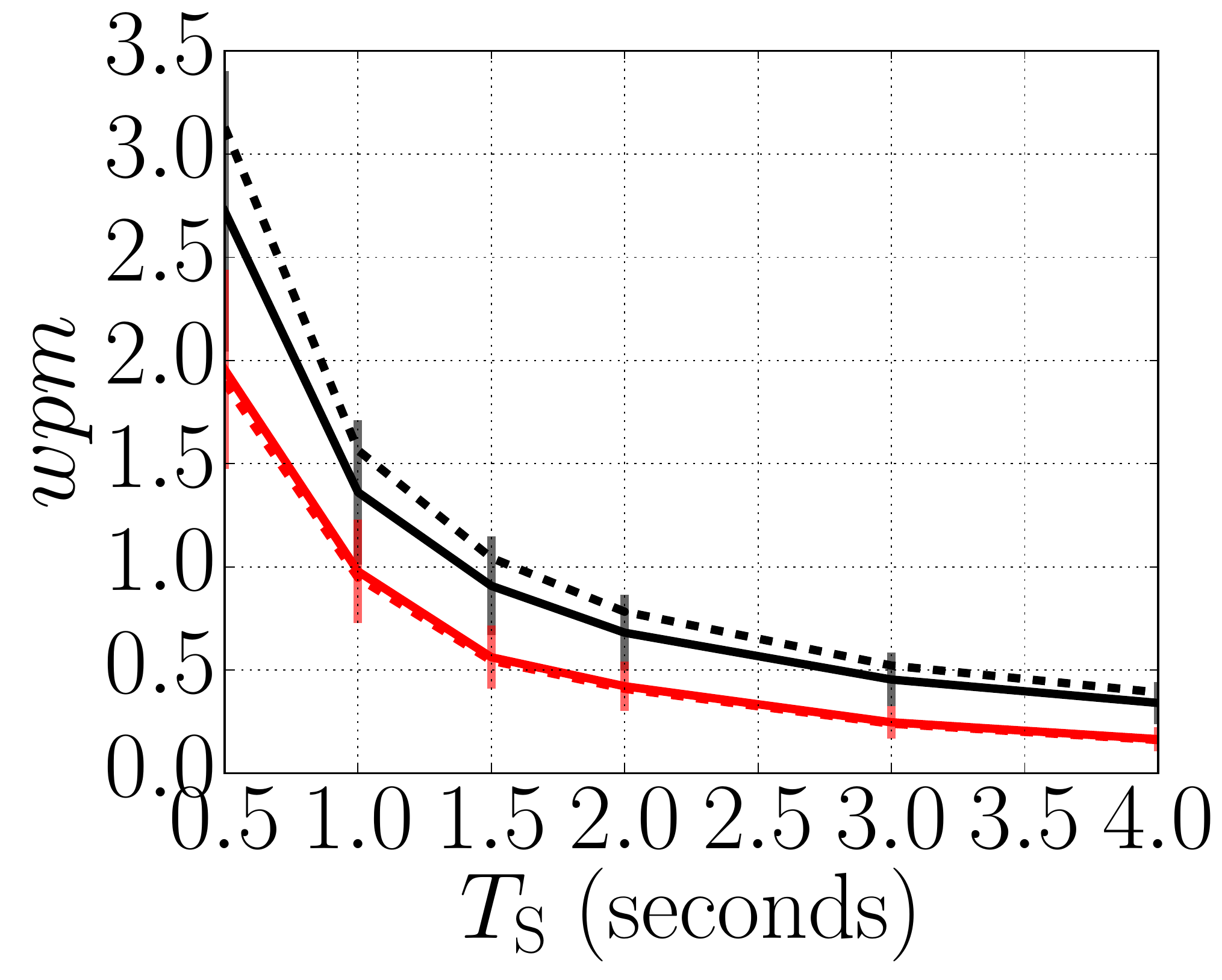} \end{minipage}
\\
\begin{minipage}{4.3cm}{(a)} \end{minipage} & 
\begin{minipage}{4.3cm}{(b)} \end{minipage}  \\ 
  
\end{tabular}
\caption{ Comparing the slow-scan (black) and fast-scan (red) methods for $\sigma > 0, \lambda=0, U=2, E=2, \kappa=10$, $\Delta=0$,  
 $T_\mathrm{S}^{\mathrm{FAST}}$ =  $T_\mathrm{S}$. The scanning delays are increased from \nu{500}{ms}, and in (a) $\sigma=\frac{T_\mathrm{S}}{6}$, 
whereas in (b) $\sigma=\frac{T_\mathrm{S}}{4}$. That is, the probability of error (clicking in another cell than $v$ when aiming for it) is 5\% in (a) and less than 1\% 
in (b). } 
\label{fig:grid2_std_dev}
\end{figure}

\figref{fig:grid2_std_dev} show the result if the standard deviation is linearly increased. This is done by making it a constant fraction of an increasing slow-scan delay. In this case the slow- and fast- method have the same scanning delay at all cells to achieve the same probability of error. Increasing the probability of error to   5\% instead of less than 1\% has a significant effect on the error bars (as more erroneous characters are selected that have to be corrected). 

\figref{fig:grid2_std_dev} is indicative of how to make decisions based on the click-timing distribution: as mentioned before $q(v'v)$ can accommodate any distribution that is continuous in time, e.g., a bimodal distribution. If the integral in \eqref{eq:overlap} can not be computed in a parametric form, it can be computed numerically (especially because there is only one dimension).
 By measuring the click-timing distribution (even if it's a rough estimate), the scanning delay can be set such that $(1-q_{v'v}) < p_{\mathrm{max}}$.
 Perhaps, if a care-worker assists the user, the system can allow more erroneous character selections than usual. However, if the system is fully automatic 
 $p_{\mathrm{max}}$ will probably have to be very small.

\subsection{Word Prediction}

Word prediction complements our current approach. However, predictions are user-dependent and not always
suitable. Users with long response times often have visual
problems and word predictions are difficult to use efficiently in standard scanning systems
when a user is visually impaired.​ ​ This is supported by the Grid2 manual, which suggests
excluding word predictions in cases of long scanning delays.

 The Markov chain in our model can, however, be extended to include word predictions, if necessary. 
%Care should be taken not 
%to create millions of states (even if the topology of the Markov chain is extremely sparse). This can happen if a state is created per 
%word in the dictionary, and from there extended to model to the number of undos and errors---the number of states will increase exponentially. 
%Since we always measure against a ground-truth word, it is more suitable to create a unique label for each word of length $m \in (1, M_\mathrm{max})$,  
%and create the appropriate transitions from there, i.e., a state must be created representing all words of, for instance, length~10 that can be selected instead 
%of the ground-truth word. This will reduce the state space substantially.

\subsection{Fast-Scan versus Slow-Scan Conclusions}

The fast-scan method requires a minor adjustment to a standard scanning system.
We have shown there is a trade-off that depends on the click-timing distribution: if the distribution has a narrow peak, but an offset, 
the fast-scan method is better to use than the slow-scan method. 
This represents a user who takes while to respond, but who responds with a high degree of consistency.

If the offset is small, but the distribution is wide, the slow-scan method is faster than the fast-scan method (for the same probability of error). 
Neither of the methods can make effective use of~$\sigma$ to 
improve precision, as neither of the methods has the ability to rescale~$\sigma$ in any other way than to increase the scanning delay.  

It is therefore necessary to measure $\Delta$ and $\sigma$ and simulate their full distribution to decided whether to use the fast- or slow scan method.

The probability of error is reduced when increasing the scanning delay linearly according to~$\sigma$ in both methods. This idea can be generalized to any distribution---after measuring the click-timing distribution, it is possible to decide on an acceptable probability of error.  
A standard scanning system can not make use of the full distribution to decrease the probability of error, other than increasing the scanning delay.

%Our simulation model can be used to generate expected baseline results. To achieve this, a user's click-timing distribution needs to be estimated using for example %%kernel-density estimation techniques to estimate the distribution. The data for estimation can be obtained by recruiting a participant and instructing the participant $to click when they hear particular letters while listening to an audio stream of letters. After a few clicks it is possible to get a rough idea of the click-timing distribution, the false-positive and false-negative rate. The distribution can then be used in our simulation model to get a baseline result of what to expect. If for example an excessive scanning delay is required to ensure an acceptable probability of error (which depends on the full click-timing distribution's range), it might be better to spend the effort on an alternative interface.

One can take the suggested fast-scanning method a step further by asking the following questions: Can the long scanning delay be deferred to the end of a letter or word (i.e., after two or more group scans)?  The layout will definitely have to change, otherwise certain letters like an ``a" will always be more difficult to select compared to some other letters.  
Including false- positive and negatives, will make the effect of the noise  prevalent. Can the alphabet be represented to the user in an alternative way to better determine what the user's intentions were?  

 Nomon \cite{nomon} answers many of these questions.
Our current research extends Nomon \cite{nomon} by deferring the scanning delay to the end of a word, and to explicitly include a model for the false- positives and negatives as part of the text entry method. We are focussing on techniques to assist users with visual impairments. We aim to model the same noise sources defined in this paper, isolate them, and compare what the effect is on the interface. By casting a scanning system into a probabilistic framework we can directly compare against the slow-scan methods via simulation.

\section{Conclusions}

In this paper we cast a widely used standard scanning system, which is not probabilistic, into a probabilistic framework.
 We identify noise sources which occur frequently in practice and which cause the most prevalent communication bottlenecks. We model these noise sources using probability distributions. We provide a framework to test the effect of these noise sources, and show that this enables a direct comparison 
to other scanning methods that support probability distributions (e.g., the proposed fast-scan method in this article).

We developed a Markov chain to model user interactions with a scanning system.
From a theoretical viewpoint, the Markov chain has been constructed to reflect usage of
at least one commercial standard scanning system (Grid2) exactly. All the settings were based on the
reference manual and the trial software for Grid2. Different noise distributions can
be plugged into the Markov chain as described above, without loss of generality.  

A few minor adjustments has been made to the simple interface suggested by the Grid2 manual to enable faster letter selections in audio mode. 
The software prototype can be found here~\cite{ticker2015}. For validation purposes, a single user 
has been trained to mimic the simulation by injecting synthetic noise into the system while the user was typing. The simulation was then compared against the user's results.

In practice, we hope that scanning system designers will take on the idea of measuring a user's click-timing distribution (in the form of e.g., a histogram), 
before asking the user to learn a new interface. It is then possible to plug this data into various simulations that support probability distributions. 
This might reduce development time
significantly, and might also lead to new design insights. At the very least, it is possible to choose the scanning delay to match an acceptable probability of error.   
 
 \section{Acknowledgements}

 This work was  funded by the Aegis EU project, Gatsby Charitable Foundation, and by Emli-Mari and her husband Robert Fanner.  
We also thank Robert Fanner for his help with the software. We thank Grid2 for providing us with free software to test, and 
the  Firelight Technologies for the free usage of the excellent audio library ``fmod".

% REFERENCES FORMAT
% References must be the same font size as other body text.
\bibliographystyle{IEEEtran}
\bibliography{references}

\begin{tabular}{c}

\begin{minipage}{7.5cm}
\begin{IEEEbiography}[{\imagetop{\includegraphics[width=1in,height=1.25in,clip,keepaspectratio]{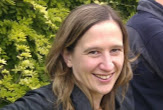}}}] 
{Emli-Mari Nel}
% graduated from the University
%of Stellenbosch, South Africa with the B-Eng degree (electronic engineering) in 2001 and the PhD (electronic engineering) in 2005.
 %She worked at Oxford Metrics Group (in Oxford) as a computer vision researcher until 2009, after which she 
 was a postdoctoral researcher with Prof David MacKay  at the Cambridge Inference Group from 2009 to
 2014. She currently works at Empiric Capital, a hedgefund in the Western Cape, South Africa, and also works as a freelance machine learning researcher.
%In her spare time she works on accessibility applications.
% She is a proud mother of two boys and is interested in Bayesian inference, machine Learning and computer vision.   
 \end{IEEEbiography}
\end{minipage}

\\
\begin{minipage}{7.5cm}
%\begin{IEEEbiographynophoto} 
\begin{IEEEbiography}[{\includegraphics[width=1in,height=1.25in,clip,keepaspectratio]{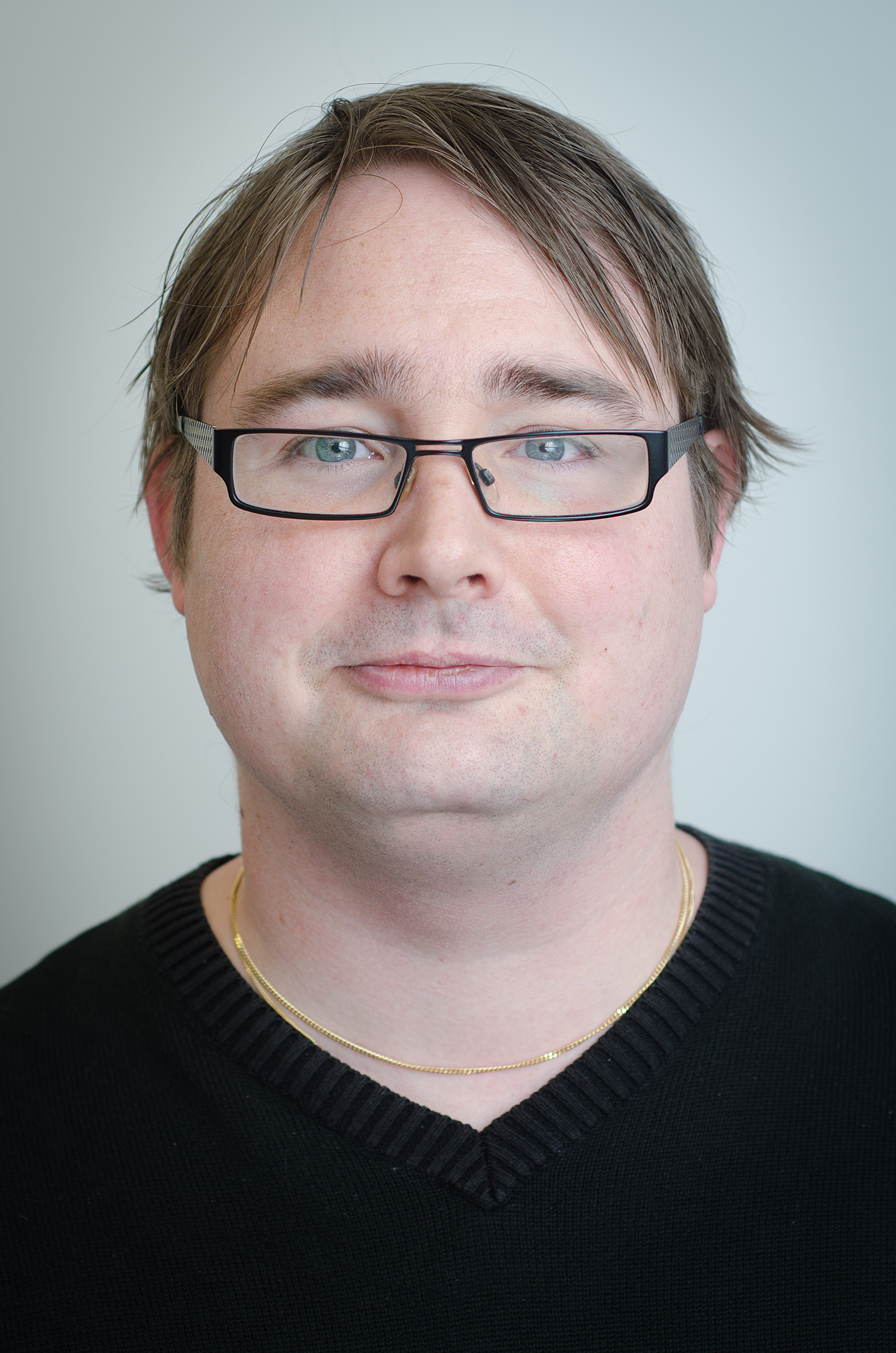}}] 
{Per Ola Kristensson} is a University Reader at the Department of Engineering, University of Cambridge 
and Fellow of Trinity College, Cambridge.
%He is an Honorary Associate
%Professor (Docent) in Computer and Systems Science at Stockholm
%University, Sweden and an Honorary Reader at the University of St
%Andrews. 
In 2013 he was recognised as an Innovator Under 35 (TR35) by
MIT Technology Review and appointed a Member of the Royal Society of
Edinburgh Young Academy of Scotland. In 2014 he won the ACM User
Interface Software and Technology (UIST) Lasting Impact Award and the
Royal Society of Edinburgh Early Career Prize in Physical Sciences,
the Sir Thomas Makdougall Brisbane Medal. 
%He is an Associate Editor of
%ACM Transactions on Intelligent Interactive Systems and the
%International Journal of Human-Computer Studies.
\end{IEEEbiography}
\end{minipage}
\\

%\begin{minipage}{7.5cm}
%\begin{IEEEbiography}[{\includegraphics[width=1in,height=1.25in,clip,keepaspectratio]{Mick}}] 
% {Mick Donegan} has many years of practice-based work as a teacher and an Assistive Technology specialist. He was Deputy Head of Wilson Stuart %Special School in Birmingham and Deputy Director of the ACE Centre, Oxford. He is an Associate Senior Research Fellow at SMARTlab, University College, Dublin, and is Adjunct Professor in the Department of Assistive Design at OCAD University, Ontario.
%\end{IEEEbiography}
%\end{minipage}
%\\

%\begin{IEEEbiographynophoto} 
\begin{minipage}{7.5cm}
\begin{IEEEbiography}[{\includegraphics[width=1in,height=1.25in,clip,keepaspectratio]{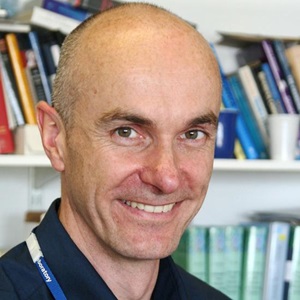}}] 
{David MacKay} (22 April 1967 --- 14 April 2016) obtained his PhD in 1992 at Caltech as a Fulbright Scholar, where his supervisor was John Hopfield. He was Chief Scientific Adviser to the UK Department of Energy and Climate Change (DECC) 
from 2009 to 2014. He was elected a Fellow of the Royal Society (FRS) in 2009. He was Regius Professor of Engineering at the University of Cambridge from 2013 to 2016.  
In 2016 he was appointed a Knight Bachelor.
He wrote two books: Sustainable Energy – Without the Hot Air (which sold 40,000 copies and has been downloaded nearly half a million times); and Information Theory, Inference, and Learning Algorithms. His contributions in machine learning and information theory include the development of Bayesian methods for neural networks, the rediscovery (with Radford M. Neal) of low-density parity-check codes, and the invention of Dasher, a hands-free keyboard.
\end{IEEEbiography}
\end{minipage}

\end{tabular}

\end{document}